\documentclass[12pt,preprint]{aastex}
\usepackage{multirow,amsthm,amsmath,amssymb,mathrsfs,ulem}
\usepackage{graphicx,subfigure,verbatim,color,soul,lineno}

\def\aap{Astron.\ Astrophys.\ }

\def\apjl{Astrophys.\ J.\ Lett.\ }
\def\apjs{Astrophys.\ J.\ Supp.\ }

\graphicspath{{fig/}}

\begin{document}

\title{Possible evidence for extended X-ray emission surrounding PSR B0656+14 with eROSITA}

\author{Shu Niu$^{1}$, Qiang Yuan$^{1,2}$, Shui-Nai Zhang$^{1,2}$, Lei Lei$^{1,2}$, Li Ji$^{1,2}$, Yi-Zhong Fan$^{1,2}$}

\affil{$^1$Key Laboratory of Dark Matter and Space Astronomy, Purple Mountain Observatory, Chinese Academy of Sciences, Nanjing 210023, China;
yuanq@pmo.ac.cn, snzhang@pmo.ac.cn\\
$^2$School of Astronomy and Space Science, University of Science and Technology of China, Hefei 230026, China\\
}


\begin{abstract}

Extended very-high-energy $\gamma$-ray emission from middle-aged pulsars as revealed recently
by several groundbased $\gamma$-ray experiments has strong implication on the transport of
high-energy particles in the interstellar medium surrounding those pulsars. The $\gamma$-ray
emission is widely believed to be produced by high-energy electrons and positrons accelerated
by the pulsar wind nebulae when scattering off the interstellar radiation field via the inverse
Compton process. Consequently, multiwavelength counterparts of the $\gamma$-ray halos are expected 
to be present, which have not yet been detected. In this work we report the possible detection 
of extended X-ray emission from a $\sim 0.2\degr$ radius region around PSR B0656+14 with eROSITA.
In spite that there are uncertainties of the on-orbit point spread function of the pointing mode, 
the radial profile of PSR B0656+14 is found to be broader than that of a star at similar
observational conditions, indicating that emission is possibly from the expected extended halo
around the pulsar. The spectrum of the emission can be described by a power-law function with 
an index of $\sim3.7$. Its surface brightness declines with radius faster than the prediction of 
the particle diffusion and synchrotron radiation in a uniform magnetic field, suggesting the 
existence of a radial gradient of the magnetic field strength as $\sim r^{-1}$. The magnetic 
field strength in the X-ray emitting region is constrained to be $4-10~\mu$G.

\end{abstract}


\section{Introduction}

The transport of Galactic cosmic rays (CRs) in the interstellar medium (ISM) is a fundamental
question of astroparticle physics. Due to the scattering with the random magnetic turbulence,
charged CRs propagate diffusively in the Milky Way \citep{1990acr..book.....B,2007ARNPS..57..285S}.
The average diffusion coefficient, usually parameterized as a function of particle's rigidity 
and assumed to be spatially uniform, can be inferred from the measurements of secondary to
primary ratios of CRs \citep{2016PhRvL.117w1102A,2022SciBu..67.2162D}. Possible spatial
variations of the diffusion coefficient are expected, due to the changes of properties of 
the ISM. However, such variations cannot be directly measured from the observations of CRs.
It has been shown that large extended $\gamma$-ray halos at very high energies (VHE)
\citep{2017Sci...358..911A,2021PhRvL.126x1103A,2017PhRvD..96j3016L,2022PhRvD.105j3013H,
2023A&A...672A.103H,2023A&A...673A.148H,2023ApJ...944L..29A}, which are expected to be 
produced by the inverse Compton scattering between high-energy electrons and positrons 
accelerated by the central pulsar wind nebulae (PWN) and the interstellar radiation field 
during the propagation process of electrons and positrons, are very powerful sources to probe 
the transport properties of particles. Particularly, a very slow diffusion is inferred from 
the morphologies of these pulsar halos, indicating that the propagation is spatially dependent
\citep{2018ApJ...863...30F,2018PhRvD..97l3008P}.

Identifying multi-wavelength counterparts of the VHE pulsar halos is particularly important in
understanding the emission mechanism of the halos and constraining the magnetic properties of the
surrounding ISM. Possible extended GeV $\gamma$-ray emitting halo around Geminga in the Fermi-LAT 
data has been reported \citep{2019PhRvD.100l3015D}, which is, however, subject to debate 
\citep{2019ApJ...878..104X}. Efforts to search for X-ray and radio counterparts have been paid, 
but no clear detection has been found yet \citep{2019ApJ...875..149L,2024A&A...683A.180K,
2024A&A...689A.326M,2024ApJ...969....9W,2024arXiv240506739H}
\footnote{The extended $\gamma$-ray source HESS J1809-193 was recently suggested to be a candidate 
pulsar halo \citep{2023A&A...672A.103H} with an X-ray halo \citep{2023ApJ...949...90L}. However, 
this region is very complicated, and the origin of the $\gamma$-ray emission is ambiguous.}.

We note that the spectral similarity between the central pulsar and the diffuse emission may result from model degeneracies, since weak components can often be fitted acceptably by multiple models. The leakage fraction beyond $4'$ is expected to vary with photon energy and should not reproduce the pulsar spectrum. Moreover, the normalization values, serving as a proxy for the source flux, differ significantly between the pulsar model and the diffuse emission\citep{2025RNAAS...9..154K}, further suggesting that the diffuse component is not dominated by PSF leakage.

In this work, we report the detection of an extended X-ray halo in a region with radius 
of 0.2\degr\ around the Monogem pulsar (PSR B0656+14, Right Ascension $=104.951117\degr$, 
Declination $=+14.239175\degr$), with the eROSITA data taken during its Performance Verification 
(PV) phase \citep{2022A&A...661A..41S}. The PWN associated with PSR B0656+14 as revealed by Chandra 
is about $15''$ \citep{2016ApJ...817..129B}, which is much smaller than the emission size found in 
this work. The most natural explanation of the extended X-ray halo is the synchrotron radiation of 
electrons and positrons which give rise to the VHE $\gamma$-ray emission. The magnetic field 
strength is thus constrained to be $4-10~\mu$G based on the multi-wavelength data.
Similar extended X-ray emission has also been reported in other compact systems, such as 
the microquasar V4641~Sgr observed with XRISM \citep{2025ApJ...978L..20S}.

\section{Data Reduction}

The eROSITA observations (OBS\_ID: 300000) of PSR B0656+14 were performed on October 14th, 2019 
in pointing mode\footnote{We did not include the data in survey mode from the eROSITA Data Release 
1 (DR1) due to very limited exposure time and quite different instrumental response.} during 
Calibration and Performance Verification (Cal-PV) phase. Due to the operation, telescope module 
(TM) 1 failed to function properly, while TM 5 and TM7 were affected by optical light leaking, 
as they do not equip an aluminium optical light filter \citep{2021A&A...647A...1P}.
To ensure optimal sensitivity in the soft X-ray band and extend the low-energy range down to 
0.2 keV, we utilized data only from TM2, TM3, TM4, and TM6. The data were processed using the 
eROSITA Science Analysis Software System (eSASS) for DR1. Periods of high background activity 
were excluded by employing the {\tt flaregti} task, yielding an effective exposure of 99.99 ks.

For spectral extraction, we defined an annular region centered on PSR B0656+14 with radii from 
4$'$ to 26$'$ to avoid central source contamination and minimize uncertainties due to edge 
effects of the instrument's field of view. Within this region, approximately 530 point sources 
were identified and excluded using the {\tt ermldet} tool (Fig.~\ref{fig:ds9}; see Appendix \ref{app:pointsource} for 
more details). This annular region was subsequently divided into 12 concentric rings, each with a 
radius increment of 2$'$, resulting in 12 spectra and their corresponding response files. When 
generating the response matrices and auxiliary response files, the parameter exttype and psftype 
of the tool {\tt srctool} were specifically set to ``tophat'' and ``none'' with assuming the the 
diffuse emission region are very extended. Each spectrum includes an instrumental background 
spectrum derived from filter wheel closed (FWC) data. Ultimately, we obtained 12 spectra spanning an 
energy range of 0.2–2 keV. These spectra are re-binned to ensure sufficient photon counts per 
channel for better statistical reliability (see Fig.~\ref{fig:spec} in Appendix \ref{app:pointsource}).

\section{Spectral fitting and results}
The signal we search for may be spatially non-uniform. However, all these 12 annular regions share 
the same diffuse background emission. Therefore, we jointly fit these 12 spectra using a common set 
of background parameters, but allowing the flux of the signal component to vary independently across 
the rings. The spectrum of the signal component is modelled as a power-law, with a constant slope 
across the 12 annuli. Its foreground absorption should be similar to that of the central point source.
\citet{2022A&A...661A..41S} analyzed the central source using various models, and estimated the HI 
column density to be about $(1\sim2.8)\times10^{20}\,\rm cm^{-2}$. We thus constrain the foreground 
column density in a similar range.

The diffuse X-ray background is theoretically a composite of several components
\citep[e.g.,][]{2023A&A...674A.195P}. 
The first is the cosmic X-ray background (CXB) predominantly originating from active galactic nuclei 
and galaxy clusters. Its spectrum is characterized by a power-law distribution with a fixed slope of
$\sim1.46$ and a fixed normalization of $\sim 11.6\, \rm photons$ $\rm cm^{-2}~s^{-1}~sr^{-1}~keV^{-1}$ 
at 1 keV. Its foreground absorption is modeled based on the HI column density of the Milky Way. 
The second background is the emission from the local hot bubble (LHB) and the solar wind charge 
exchange (SWCX). In CCD spectra, these components can be modeled using an APEC plasma model with 
a temperature of $\sim$0.1 keV, assuming solar metallicity and no foreground absorption
\citep{2023A&A...674A.195P}. The third one is the emission from the Milky Way's circum-galactic 
medium (CGM), Galactic halo, and unresolved stellar contributions. The CGM contains a hot, tenuous 
plasma, with a characteristic temperature of 0.15-0.2 keV and low metallicity. Some high-temperature 
plasma around 0.4-0.7 keV in the halo or the Galactic corona and unresolved M-dwarf stars contribute 
soft X-rays in the 0.6-2 keV band.

Consequently, an empirical fitting approach is employed, utilizing two absorbed APEC components for 
the third background emission: one with an initial temperature around 0.2 keV with free metallicity 
and the other is fixed at 0.7 keV with solar abundance. We fix the temperature and metallicity of 
the latter one due to the insufficient constraints of the data as \citet{2024A&A...690A.399Y}.
Both APEC components (CGM and Galactic halo) and CXB share the same fixed absorption HI column 
density of $1.28 \times10^{21}\,\rm cm^{-2}$ from the HI observation. In this work, we use the 
{\tt tbabs} model \citep{2000ApJ...542..914W} to estimate the ISM absorption. Finally, the adopted 
model of the background  components is {$\tt (tbabs*(powerlaw_{CXB}+apec_{CGM}+apec_{Halo})+apec_{LHB})$}.

\begin{figure}[!htbp]
\centering
\includegraphics[width=0.6\textwidth]{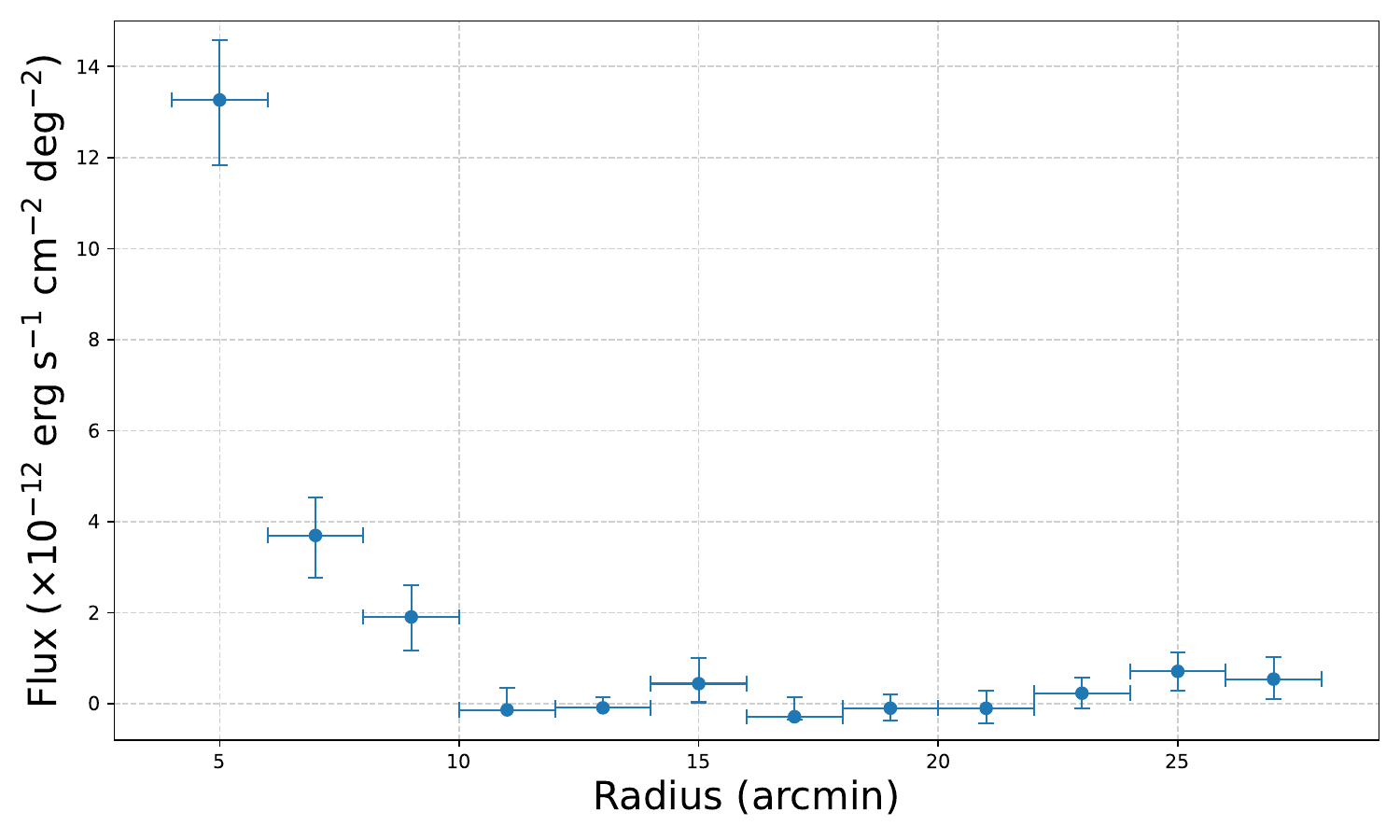}
\caption{Fluxes of excess X-rays in 0.2 - 2.0 keV band of the 12 annuli, after correction of the inefficient detection of weak point sources beyond $16'$. The background model assumed is the physically motivated three APEC plus CXB model.}
\label{fig:flux_corrected}
\end{figure}

Spectral fitting was performed using the Interactive Spectral Interpretation System
\citep[ISIS;][]{2000ASPC..216..591H}\footnote{Version 1.6.2; \url{https://space.mit.edu/cxc/isis}}. 
The fitting shows that a common spectrum power-law excess with photon index of $\sim 4$ on top of the
background is found in the data. The fluxes of the power-law excess vary across all annuli, showing a
decrease within $4'-10'$, reaching the level of zero within $10'-18'$, and rising again at the outermost
annuli beyond $18'$. The results are presented in Table~\ref{tab:result} and Fig.~\ref{fig:flux}. 
We expect that the rising behavior at large radii may be due to the contribution of weak point sources 
which are below the detection limit. As shown in Fig.~\ref{fig:source_detection}, the surface brightness 
distributions for sources with relatively high fluxes keep flat across the field of view. However, for 
weak sources, the detection efficiency starts to drop beyond $16'$. After correcting the expected
contamination from such weak sources, the surface brightness profile of the excess component becomes 
flat at large radii, as shown in Fig.~\ref{fig:flux_corrected}.

\begin{figure}[!htbp]
\centering
\includegraphics[width=0.6\textwidth]{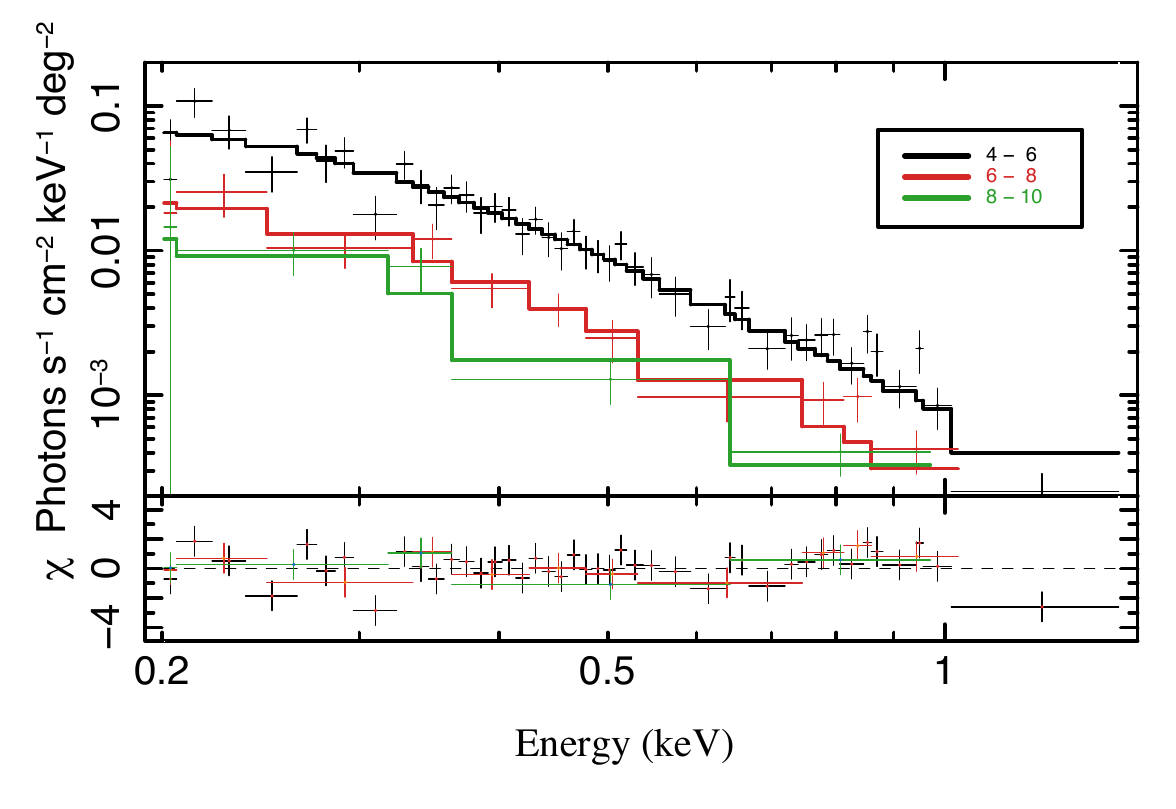}
\caption{X-ray spectra of the non-thermal component for the $4'-6'$ (black), $6'-8'$ (red), and $8'-10'$ (green) annuli. The background model is the $13'-16'$ data-driven background. The data are binned accordingly to enable that the signal-to-noise ratio (S/N) of each bin at least 3.}
\label{fig:3ring}
\end{figure}

To further verify the detected excess component in the inner three annuli, we use the spectrum of the 
$13' - 16'$ annulus as data-driven background and re-fit the spectra of the inner three annuli. In this
procedure, the instrumental background, i.e. FWC data\footnote{\url{https://erosita.mpe.mpg.de/dr1/AllSkySurveyData\_dr1/FWC\_dr1/}}, has been subtracted
individually for both the signal and background regions. In addition, the BACKSCAL keywords of the three
spectra are adjusted to the BACKSCAL columns to account for the energy-dependent differences in their 
ARFs relative to the background spectrum. The fitting results of model 
${\tt tbabs_{diffuse}*powerlaw_{diffuse}}$ for these three annuli are shown in Fig.~\ref{fig:3ring} 
and summarized in Table \ref{tab:3ring}. The results are consistent with the joint fitting of the 12 
spectra within uncertainties, further supporting the detection of an excess power-law component in the
$4'-10'$ region around PSR B0656+14.

The central pulsar is very bright in X-rays. Due to the limited point spread function (PSF) 
of the instrument, the central source may contaminate the emission in outer regions. 
The PSF model based on groundbased calibration \citep{2022A&A...661A...1B} is limited 
in accuracy at large radii. We therefore choose a pointing-mode observation of a bright 
variable star, V*~V702~CrA, to test the effect of the PSF leakage. Note that, V* V702 CrA 
was observed at an off-axis angle of $4.5'$, which is even bigger than the off-axis angle 
of PSR B0656+14 ($2.5'$). Instrumentally, a larger off-axis angle is expected to give an 
even broader PSF. As shown in Fig.~\ref{fig:sbp_comparison} of Appendix \ref{app:psf},
the radial profile of V*~V702~CrA is more concentrated than that of PSR~B0656+14 in the
0.2--2.0 keV band, supporting the presence of the diffuse emission. The fractional energy 
beyond $4'$ of V*~V702~CrA agrees well with the PSF model prediction.
We estimate that the contribution from the central pulsar to the detected power-law 
excess can be neglected in the $4'-6'$ region, with possible contamination being smaller 
than 10\% given the excess flux is about 1.4\% of the central pulsar's flux  
(see Appendix \ref{app:pointsource}). 
As a comparison, an estimated fraction of $\sim2.7\%$ of the central source is expected
beyond $4'$, according to the eROSITA survey-mode PSF \citep{2023A&A...670A.156C}. This
indicates that the PSF in the survey mode is noticeably degraded due to the instrumental 
blurring compared to that in the pointing mode around the optical axis.
These results indicate that the observed extended excess around PSR~B0656+14 is not likely 
to be an artifact of the PSF. While a well-calibrated PSF remains crucial for such analyses, 
future deeper observations and improved characterization of the eROSITA pointing PSF will 
help confirm this conclusion.

\begin{table*}[!htbp]
\caption{Fitting results of the non-thermal diffuse emission in three annuli, assuming the $13'-16'$ data-driven background.}
\label{tab:3ring}
\centering
 \begin{tabular}{c | c c c c}
 \hline
 region & $N_{\rm H}$&$\Gamma$ &  Flux$_{0.2 - 2.0~\rm keV}$
 & $\chi^2$/dof \\
 (arcmin) & ($10^{20}~{\rm cm^{-2}}$) & & ($10^{-12}~{\rm erg~cm}^{-2}~{\rm s}^{-1}~{\rm deg}^{-2}$) & \\
 \hline\hline
 $4 - 6$ &\multirow{3}{*}{$2.08^{+1.36}_{-1.16}$} & \multirow{3}{*}{$3.69^{+0.40}_{-0.36}$} & $11.20^{+4.06}_{-2.67}$ & \multirow{3}{*}{35.93/54}\\
 $6 - 8$ & &  &  $3.64^{+1.47}_{-1.01}$ & \\
 $8 - 10$ &  &  &  $2.05^{+1.01}_{-0.73}$ & \\
 \hline
 \end{tabular}
 \tablecomments{Errors are at 90\% confidence level. Columns from left to right are: annulus region, HI column density, photon power-law index, flux in 0.2 - 2.0 keV, and fitting $\chi^2$ value divided by the number of degree-of-freedom (dof).}
\end{table*}

Using the data from eRASS:4 (stacked data from the first four consecutive all-sky surveys),
\citet{2024A&A...683A.180K} searched for extended X-ray emission from 5 middle-aged pulsars, and 
found no significant diffuse emission. The exposure in their analysis is about 500 s for PSR 
B0656+14, which is perhaps too short to give a sensitive detection. The flux upper limit obtained 
is about $1.89\times10^{-12}$ erg s$^{-1}$ cm$^{-2}$ deg$^{-2}$ in the $0.5\degr-1.0\degr$ 
region and 0.5 - 2.0 keV band, assuming $\Gamma=2$. Their region of interest is beyond the signal 
region in this work, and the comparison is not direct. Furthermore, the sensitivity of point source
detection is not as high as in this analysis, and the unresolved source population may result in 
a too high background estimate, which affects the detection of possible excess. 
Using the XMM-Newton observations, an upper limit of $3.3\times10^{-14}$ erg s$^{-1}$ cm$^{-2}$ 
in the 0.5 - 5.0 keV band within $400''$ to $750''$ annulus was derived \citep{2024ApJ...969....9W}. 
In the same sky region and energy band, we obtain a flux of $2.70^{+0.91}_{-0.69}\times10^{-14}$ 
erg s$^{-1}$ cm$^{-2}$, which is consistent with the XMM-Newton limit.
These results confirm the robustness of the detected diffuse component and motivate 
further modeling of its physical origin, as discussed below.

\section{Interpretation}

The non-thermal X-ray excess detected by eROSITA shows a radially decline feature,
strongly supporting its association with the central pulsar/PWN instead of the
background emission associated with the supernova remnant or other sources. 
The emission also extends far beyond the PWN of PSR B0656+14. These features 
indicate that it is very likely to be the counterpart of the VHE gamma-ray halo. 
We thus use a halo model to describe its X-ray to VHE gamma-ray emission.

High-energy electrons and positrons are accelerated by the PWN, which are then
injected into the ISM. These particles propagate in the surrounding ISM, and
interact with the background radiation field and magnetic field, producing 
gamma-ray emission via the inverse Compton scattering (ICS) process and X-ray
emission via the synchrotron process. The injection spectrum is assumed to be a
super-exponentially cutoff power-law function, $Q(E_e)=Q_0 E_e^{-\alpha}\exp
[-(E_e/E_{\rm cut})^{\beta}]$, and the injection rate is assumed to be
proportional to the spin-down luminosity of the pulsar, $L_{\rm sd}(t) 
=L_0(1+t/\tau)^{-2}$, where $\tau\approx 10^4$ yr is a characteristic time
scale of the evolution of the pulsar spin-down luminosity. The diffusion
coefficient is assumed to be spatially uniform, energy-dependent, as
$D(E_e)=D_0(E_e/{\rm GeV})^{\delta}$. Here we do not introduce the non-uniform
diffusion \citep{2018ApJ...863...30F} or anisotropic diffusion 
\citep{2019PhRvL.123v1103L} since we focus on relatively small spatial scales 
(within 10\degr\ around the pulsar). The diffusion coefficient is adopted
to be $D(100~{\rm TeV})=4.5\times10^{27}$ cm$^2$~s$^{-1}$ as inferred from
VHE gamma-ray observations \citep{2017Sci...358..911A}, and $\delta=1/3$ is
assumed. Electrons and positrons experience strong energy losses during the
propagation. We include the synchrotron cooling and the ICS cooling with three
gray-body components of the background radiation field \citep{2017Sci...358..911A}:
the cosmic microwave background with temperature $T=2.73$ K and energy density
$u=0.26$ eV~cm$^{-3}$, an infrared background with $T=20$ K and $u=0.3$
eV~cm$^{-3}$, and an optical background with $T=5000$ K and $u=0.3$ eV~cm$^{-3}$.
The ICS cooling rate is calculated following the new formalism of
\citet{2021ChPhL..38c9801F}.

\begin{figure}[!htbp]
\centering
\includegraphics[width=0.48\textwidth]{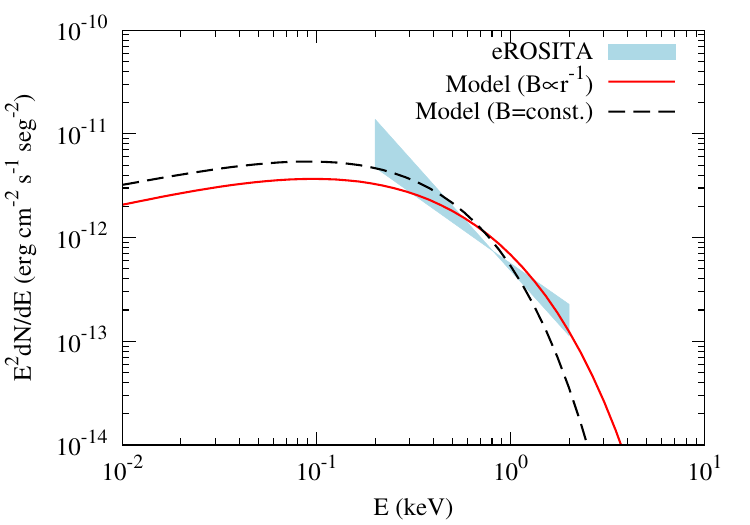}
\includegraphics[width=0.48\textwidth]{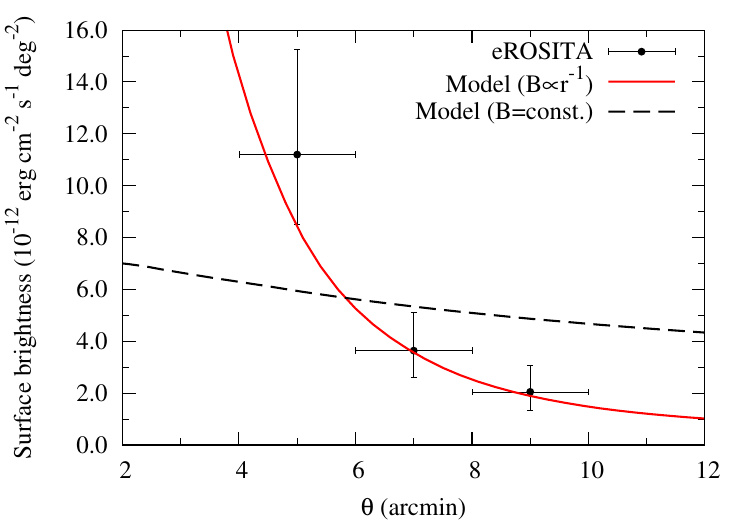}
\caption{Model predicted X-ray spectrum (left) and surface brightness profile (right) of the non-thermal component, compared with the eROSITA data. Dashed lines are for the constant magnetic field assumption, while solid lines correspond to the assumption of $B\propto r^{-1}$ in the inner pc region around the pulsar.}
\label{fig:model_xray}
\end{figure}

The injection spectral parameters are tuned to describe both the HAWC gamma-ray
data and the eROSITA X-ray data. We find that for $\alpha=1.5$, $E_{\rm cut}=35$
TeV, and a very sharp cutoff with $\beta=10$, the measured results can be
reproduced, as shown in Fig.~\ref{fig:model_xray} and Fig.~\ref{fig:model_gamma}.
The conversion efficiency from pulsar spin-down power to the electron/positron
power is about 20\%. However, we find that the spectral shapes measured by HAWC 
and eROSITA are hard to be explained well simultaneously. The eROSITA spectrum 
is very soft, with $\Gamma\approx 3.7$. To account for such a soft spectrum, we
require a sharp cutoff of the injection spectrum. Such an injection spectrum 
gives also a very soft gamma-ray spectrum (Fig.~\ref{fig:model_gamma}), which 
seems to be too soft compared with that measured by HAWC
\citep{2017Sci...358..911A}. Improved spectral measurements in both gamma and 
X rays are crucial to further understanding this issue.

The model predicts a relatively flat radial profile, if a constant magnetic 
field strength ($B=4.2~\mu$G) is assumed, as shown by the dashed in the right 
panel of Fig.~\ref{fig:model_xray}. The eROSITA data shows a fast decrease trend 
of the emission, different from the model prediction. It may indicate that there
is a radial gradient of the magnetic field. Assuming that\footnote{Here we 
assume that the magnetic field approaches to a constant ISM value of 2.5 $\mu$G
at large distance from the pulsar.} $B(r)={\rm max}[3.3(r/{\rm pc})^{-1},2.5]~\mu$G, 
the measured radial profile can be reproduced, as shown by the solid line in
Fig.~\ref{fig:model_xray}. 

\section{Conclusion and discussion}

Extended VHE gamma-ray halos around middle-aged pulsars have attracted great
attention in recent years, due to their broad connection with the acceleration
of energetic particles in the PWN, the propagation of these particles in the
surroundings of pulsars, as well as the properties of the ISM. The multi-wavelength 
counterparts of the gamma-ray halos have been expected to be present, which are, 
however, not detected yet in any bands other than VHE gamma rays. In this work, 
we report the first detection of such a counterpart associated with PSR B0656+14 
in the X-ray band, with 100 ks observations carried out by eROSITA. An excess 
component with a power-law spectrum (spectral index $\Gamma=3.69$) extending up 
to $10'$ away from the central pulsar has been identified in the data. 
The emission size of the excess is much more extended than the X-ray PWN of 
PSR B0656+14, indicating that it may be produced by particle escaping from the PWN. 
The fluxes of the excess emission also show fast decline with the increase of 
distance from the pulsar, which supports its association with the central object 
rather than unknown background. The pulsar halo model with a high-energy electron 
and positron cloud surrounding the pulsar/PWN can account for the X-ray and 
gamma-ray observations simultaneously. The flux of the X-ray emission and its radial 
profile give a direct measure of the magnetic field in the vicinity of PSR B0656+14, 
which we find favors the existence of a radial gradient as $B(r)\propto r^{-1}$. 

The identification of weak extended emission is subject to uncertainties of
the PSF model due to the lack of on-orbit data of the pointing observation mode.
A better understanding of the on-orbit PSF is very important in confirming the
results reported in this work.
Physically the extended X-ray emission associated with ultra-high energy $\gamma$-ray 
sources is not unexpected. For instance, \citet{2025ApJ...978L..20S} have recently 
reported the detection of X-ray diffuse emission around the PeVatron microquasar 
V4641~Sgr with XRISM observation, supporting the possibility of extended high-energy 
halos associated with different types of compact sources. Multi-wavelength
observations are important in pinning down the radiation nature of these source.

The spectrum of the X-ray excess appears softer compared with the HAWC $\gamma$-ray
observations. The simple isotropic diffusion model thus has difficulty in
simultaneously explaining both the X-ray and $\gamma$-ray spectra. Improved
measurements of the $\gamma$-ray spectrum with higher sensitivity and better
energy resolution, e.g., by LHAASO \citep{2022ChPhC..46c0001M}, will help clarify
the spectral behavior at the VHE band. From the modelling perspective, as shown
in \citet{2024ApJ...969....9W}, the anisotropic diffusion scenario may produce a
softer X-ray spectrum than the isotropic case while keeping similar TeV
$\gamma$-ray spectra. Future high-resolution X-ray missions such as Athena
will also provide crucial constraints on the diffuse emission.

\section*{Acknowledgements}
This work is supported by the National Natural Science Foundation of China 
(Nos. 12220101003, 12321003, 12273112) and the Project for Young Scientists 
in Basic Research of Chinese Academy of Sciences (No. YSBR-061).

\appendix

\setcounter{figure}{0}
\renewcommand\thefigure{A\arabic{figure}}
\setcounter{table}{0}
\renewcommand\thetable{A\arabic{table}}

\section{Point source contamination}
\label{app:pointsource}
The central point source, PSR B0656+14, is significantly brighter than the non-thermal component 
we aim to detect. Therefore, it is crucial to accounting for any residual effects from the central 
source due to the PSF. The eROSITA PSF is complex, clearly depending on both energy and off-axis angle.
Fortunately, the PSF has been thoroughly characterized and accurately modeled using shapelet functions
\citep{2022A&A...661A...1B}, with full sets of parameters stored in the calibration files.

\begin{figure}[!htb]
\centering
\includegraphics[width=0.7\textwidth]{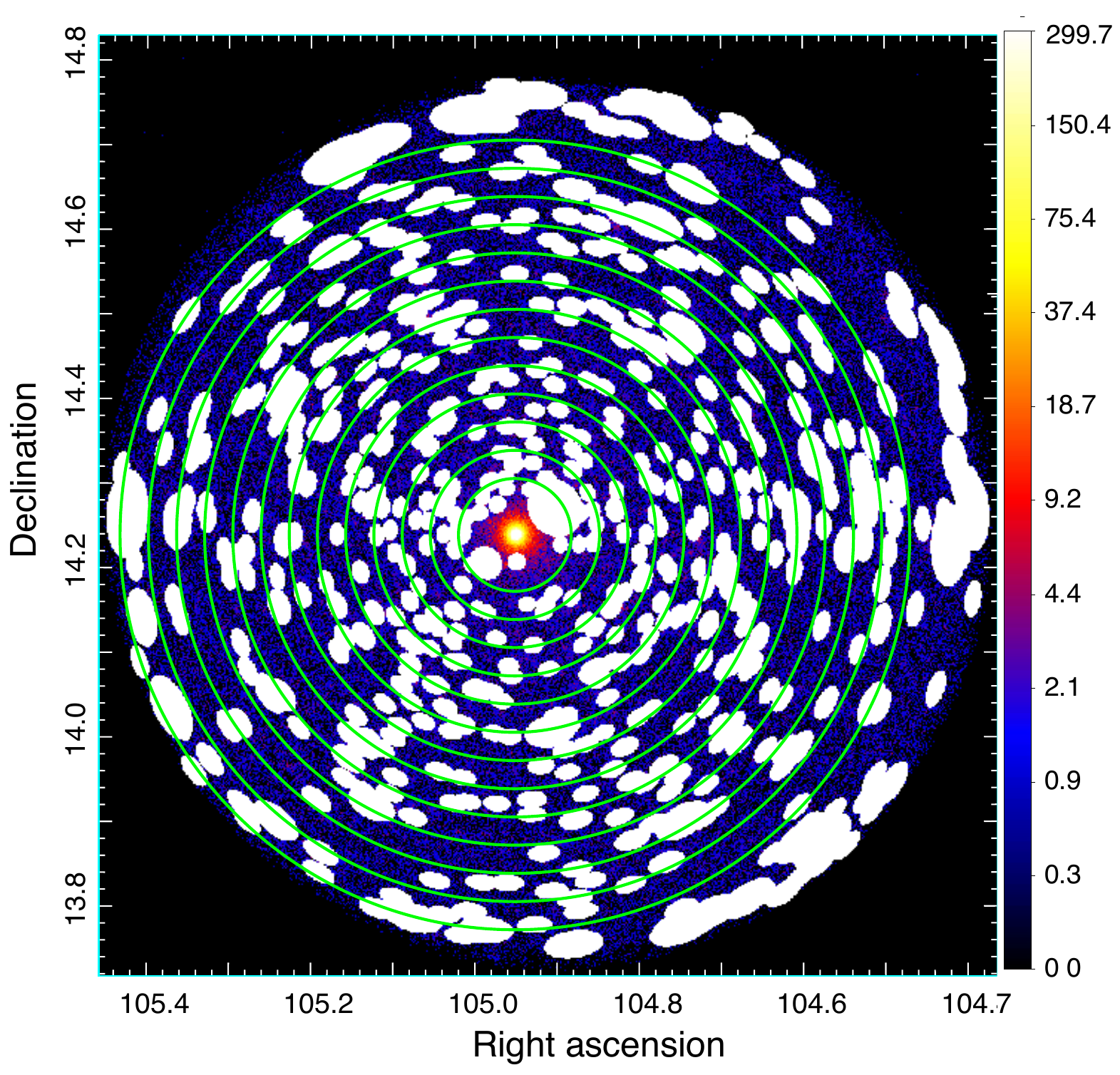}
\caption{Count map in 0.5 - 2.0 keV band of the one-degree diameter region around PSR B0656+14 
observed by eROSITA with TM2-7. White ellipses mask identified point sources out. The solid green 
circles show the boundaries of the annuli divided for spectral analysis, from 4 arcmin to 28 arcmin 
with an increment of 2 arcmin. PSR B0656+14 is shown for illustration, but also excluded in 
spectroscopic analysis.}
\label{fig:ds9}
\end{figure}

PSR B0656+14 is slightly offset ($\sim2.5'$) from the optical axis. To provide a conservative upper bound on the expected PSF leakage, we adopt a position corresponding to an off-axis of $-5'$ in the $x$ and $0'$ in the $y$ directions as a reference, where the $5'$ is the minimum spatial interval for Shapelet PSF parameters.
The Shapelet PSF images for this position from TM2, 3, 4 and 6 are combined to examine the fractions of flux
contribution at various radii. At energies of 0.3, 1.5, and 3.0 keV, the flux contributions from 
the point source in the $3'-4'$ annulus are approximately 0.3\%, 0.6\%, and 0.7\%, respectively.
Beyond 4$'$, the flux contribution become negligible according to the PSF model, with possible 
contribution $\lesssim0.1$\%. In the $4'-6'$ annulus, the flux of the non-thermal component is 
estimated to be $\sim$1.4\% of the central point source in 0.2-2.0 keV. Therefore, starting the 
innermost annulus at a radius of 4$'$ is justified, ensuring that the fitting process is not 
contaminated by the central source.

To exclude the contributions of other point sources, we utilize the {\tt ermldet} tool to mask 
them according to their PSF profiles. In the pipeline, the semi-minor axis of the source masks 
is typically around $\sim40''$. From the above PSF images, within a radius of 1$'$, the enclosed 
flux fractions are approximately 96.5\%, 94.4\%, and 94.3\% at 0.3, 1.5, and 3.0 keV, respectively.
According to the eROSITA point source catalog for this region, there are six sources whose fluxes 
are about 1/50 of that of PSR B0656+14. For these sources, we extended the mask radius to 1.2$'$ 
to ensure that any PSF residuals would not affect the measurement of the non-thermal component.
The final count map after source mask is shown in Fig.~\ref{fig:ds9}.

\begin{figure}[!htbp]
\centering
\includegraphics[width=0.32\textwidth]{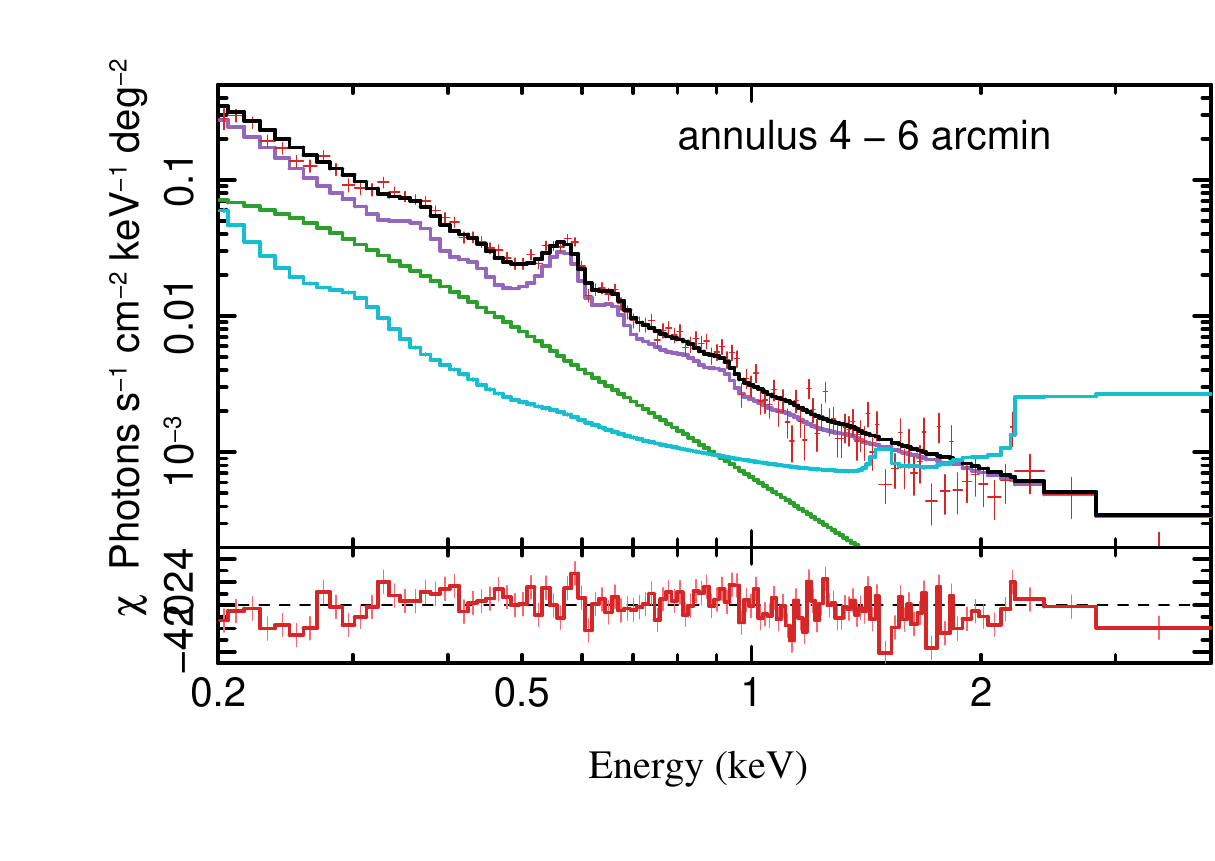}
\includegraphics[width=0.32\textwidth]{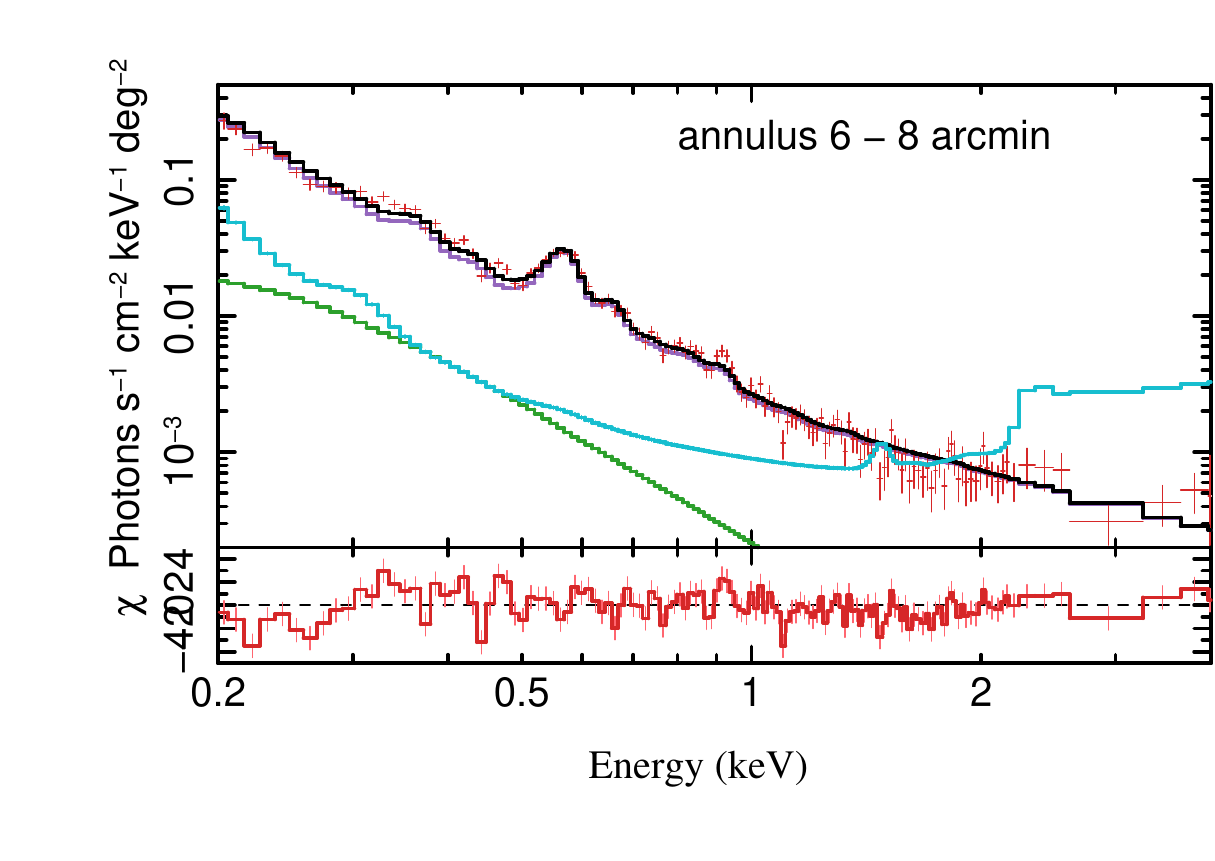}
\includegraphics[width=0.32\textwidth]{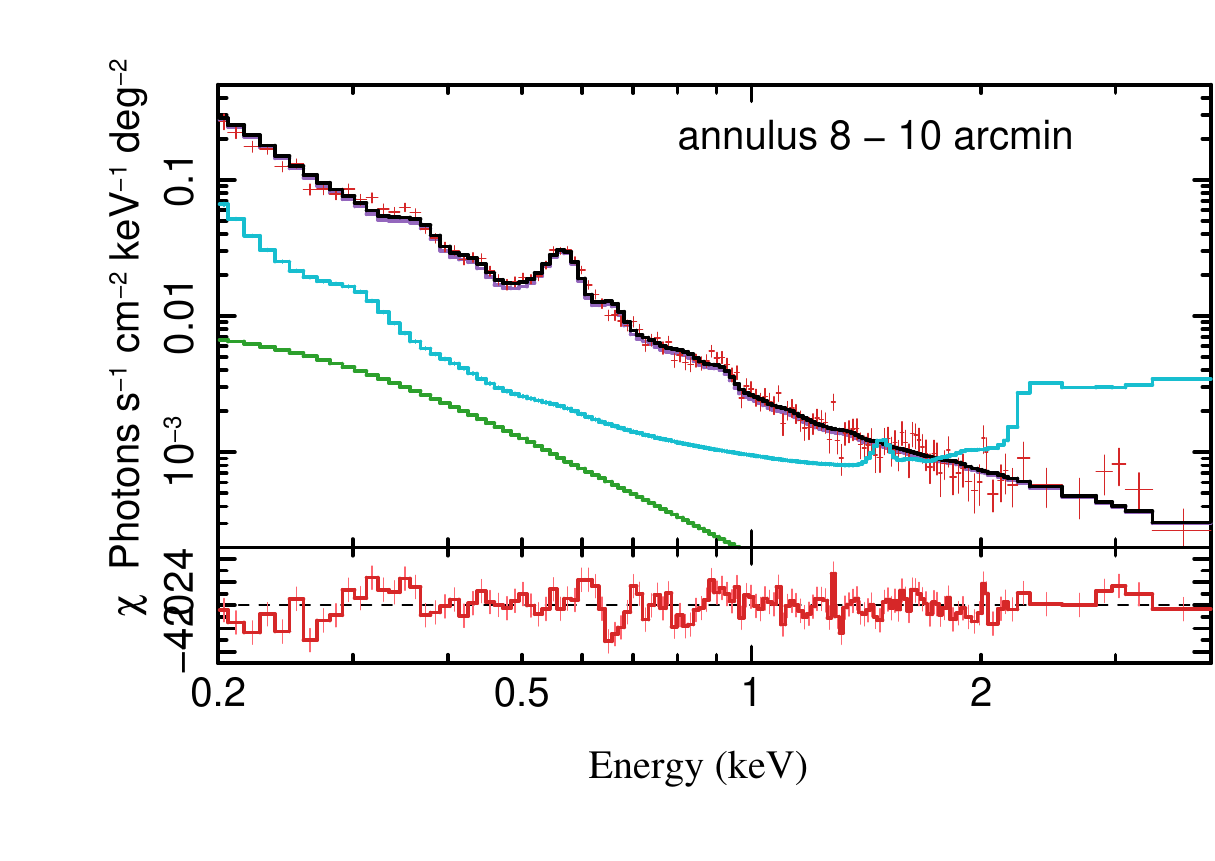}
\includegraphics[width=0.32\textwidth]{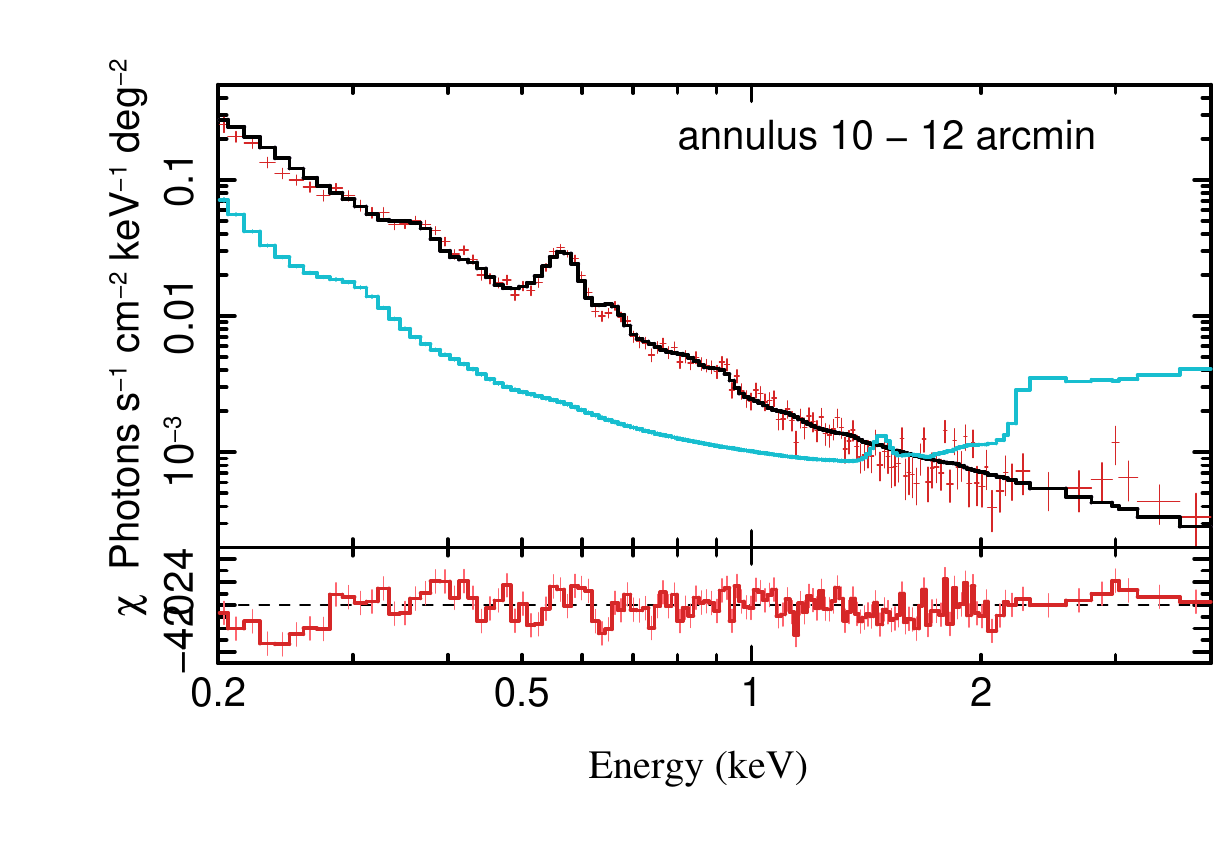}
\includegraphics[width=0.32\textwidth]{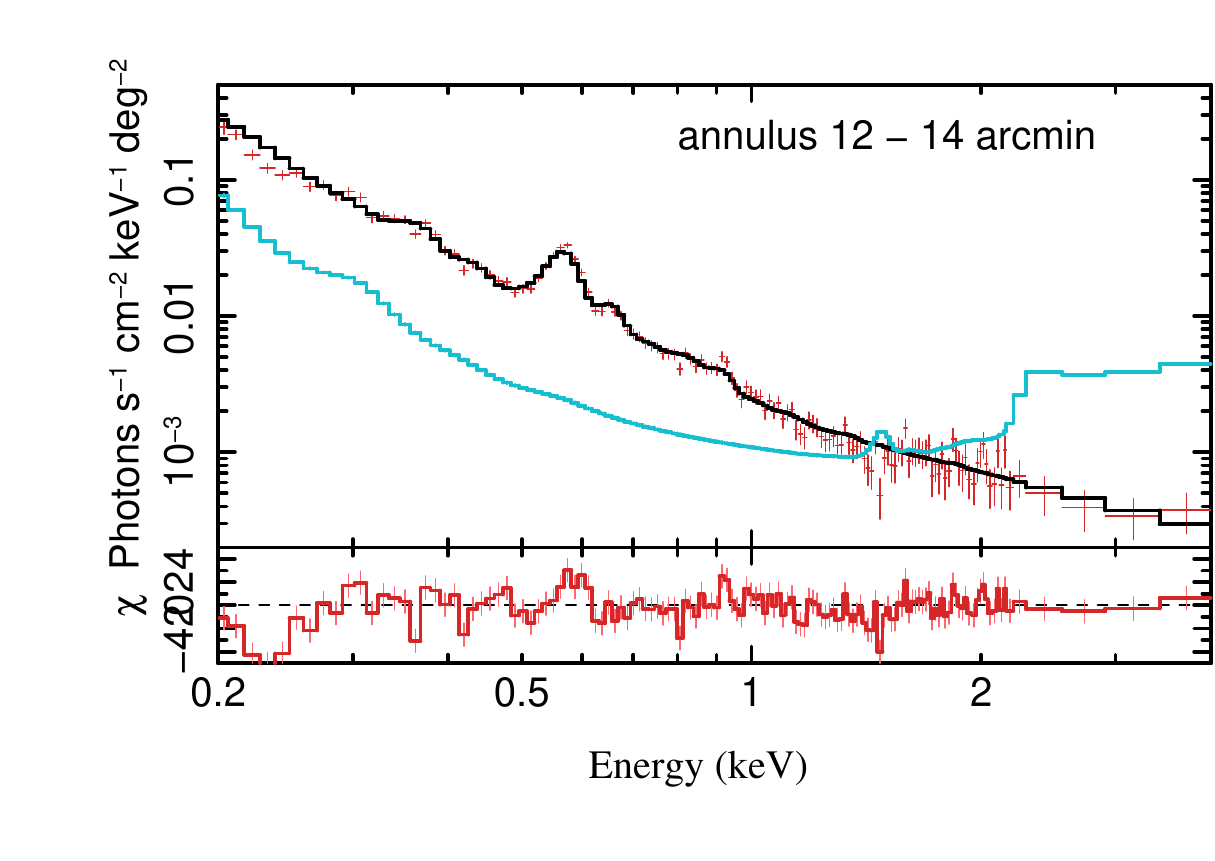}
\includegraphics[width=0.32\textwidth]{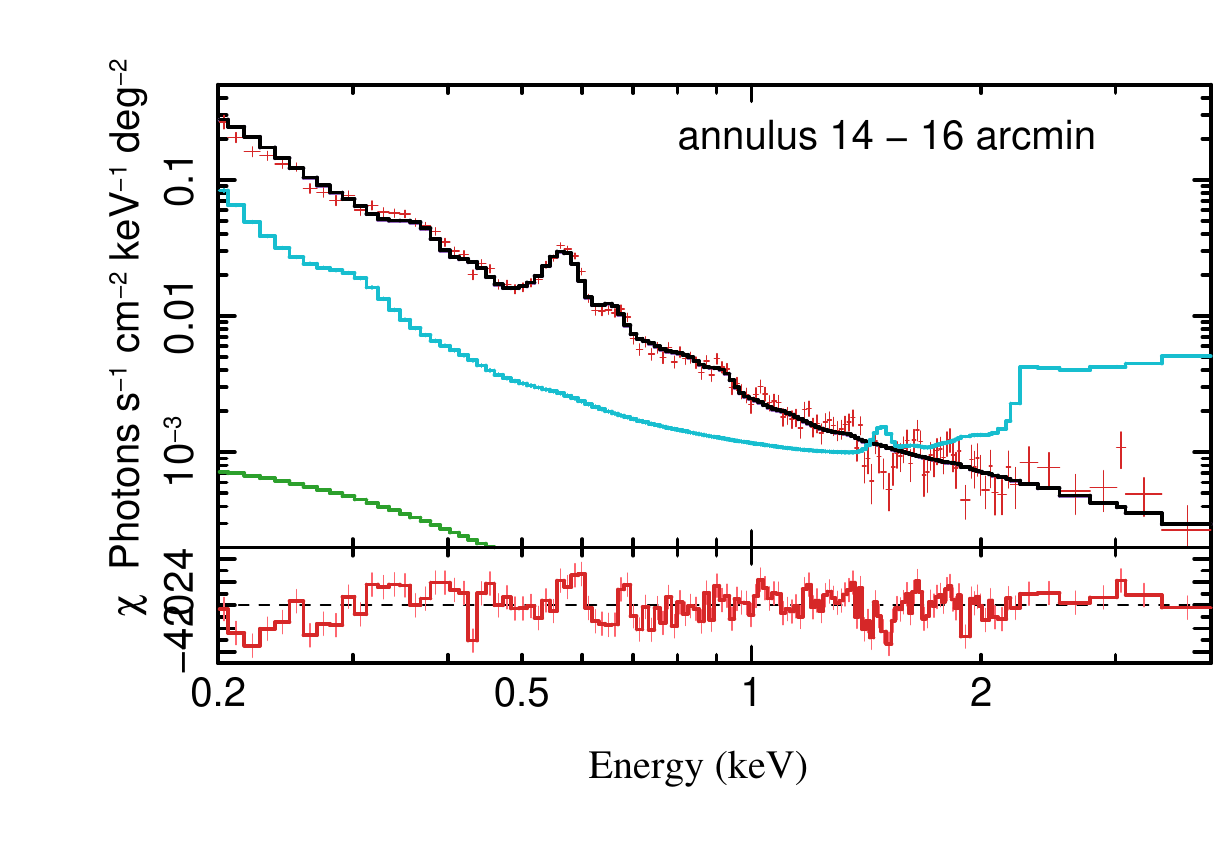}
\includegraphics[width=0.32\textwidth]{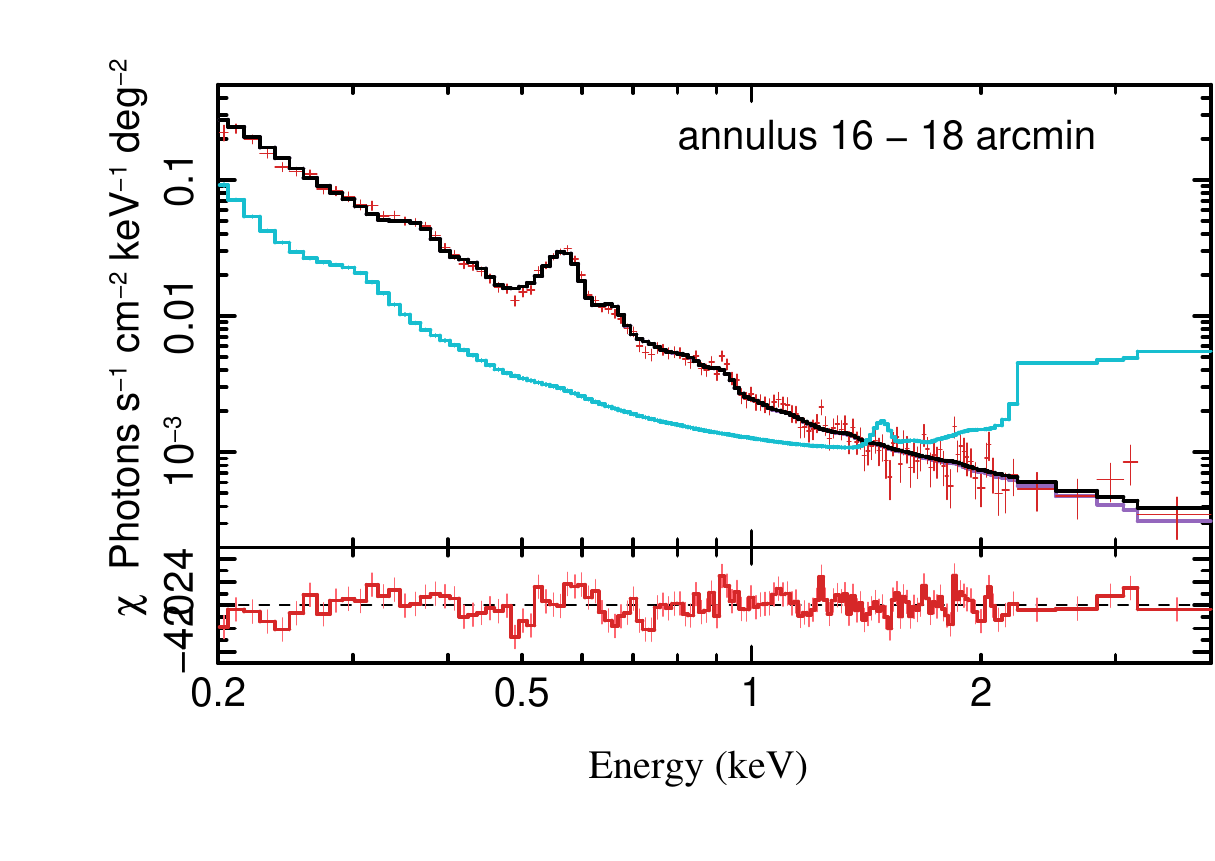}
\includegraphics[width=0.32\textwidth]{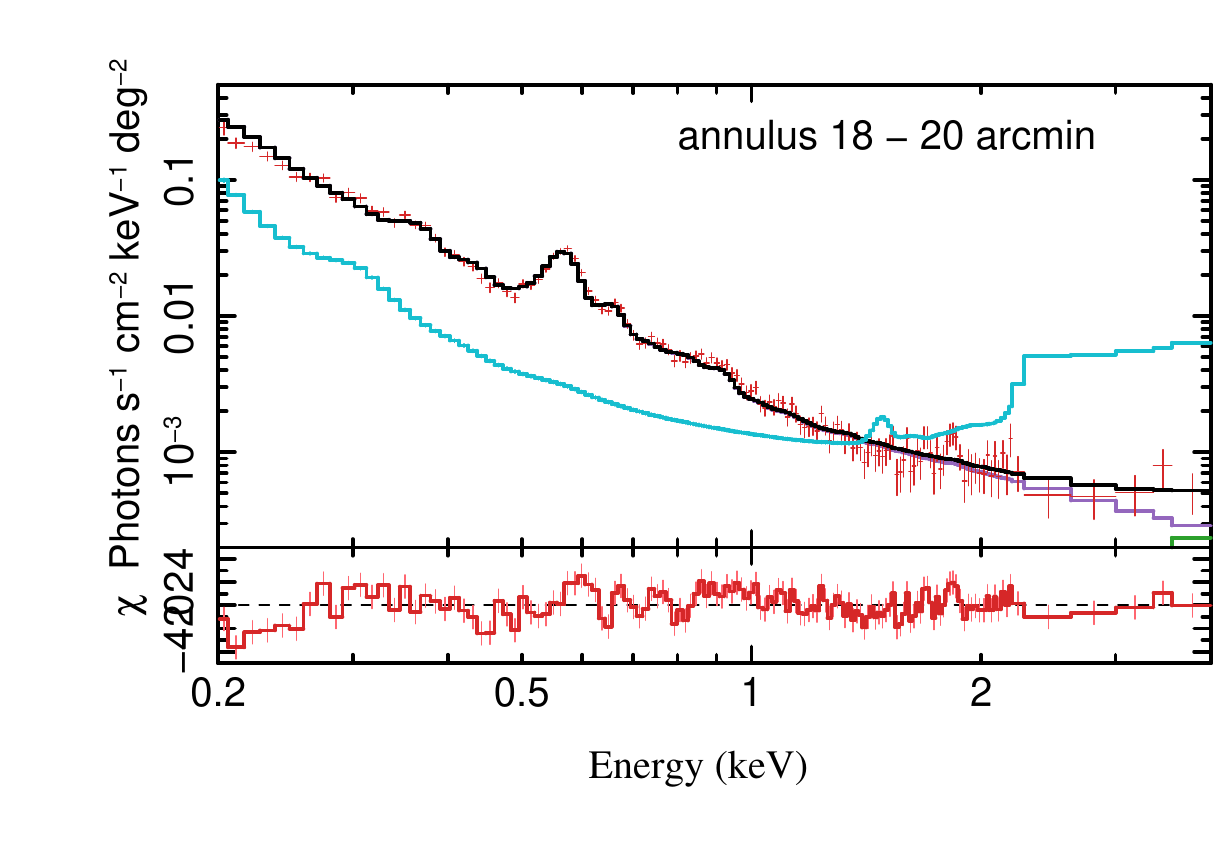}
\includegraphics[width=0.32\textwidth]{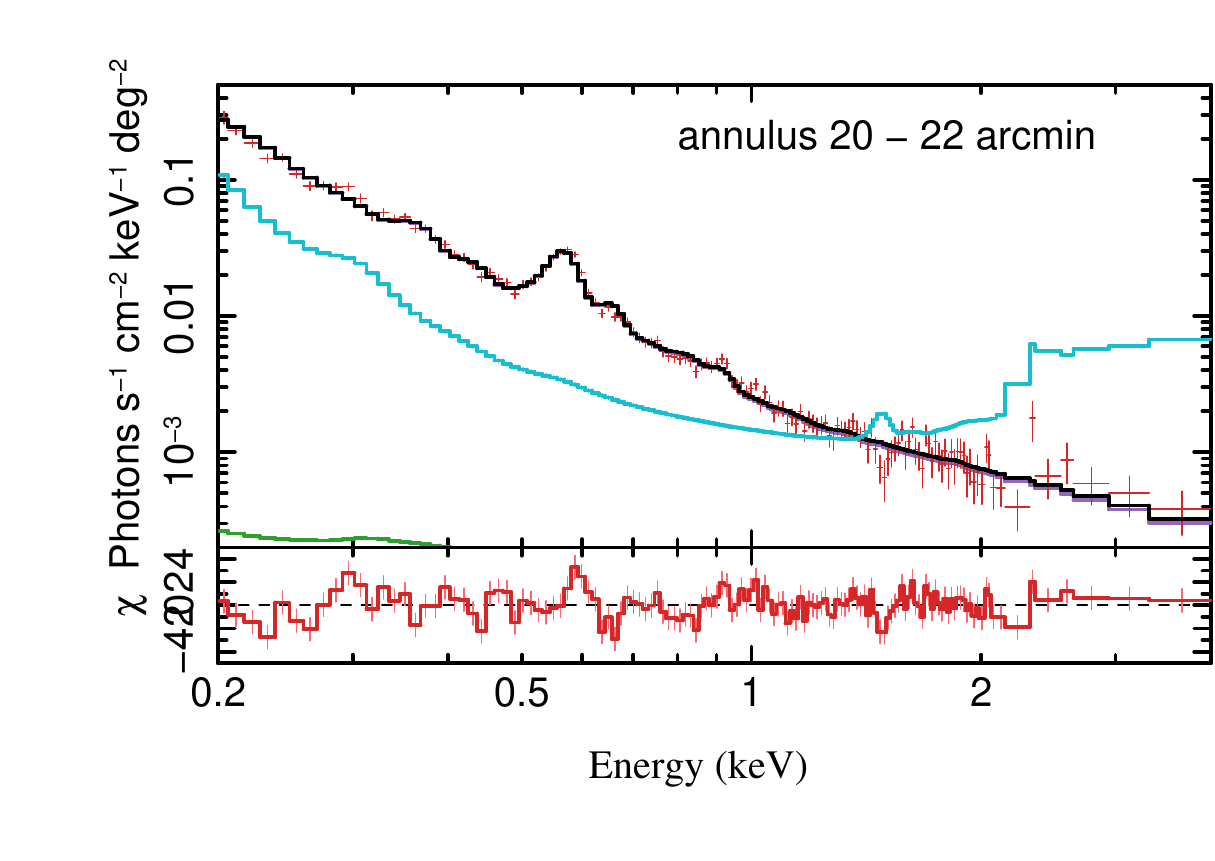}
\includegraphics[width=0.32\textwidth]{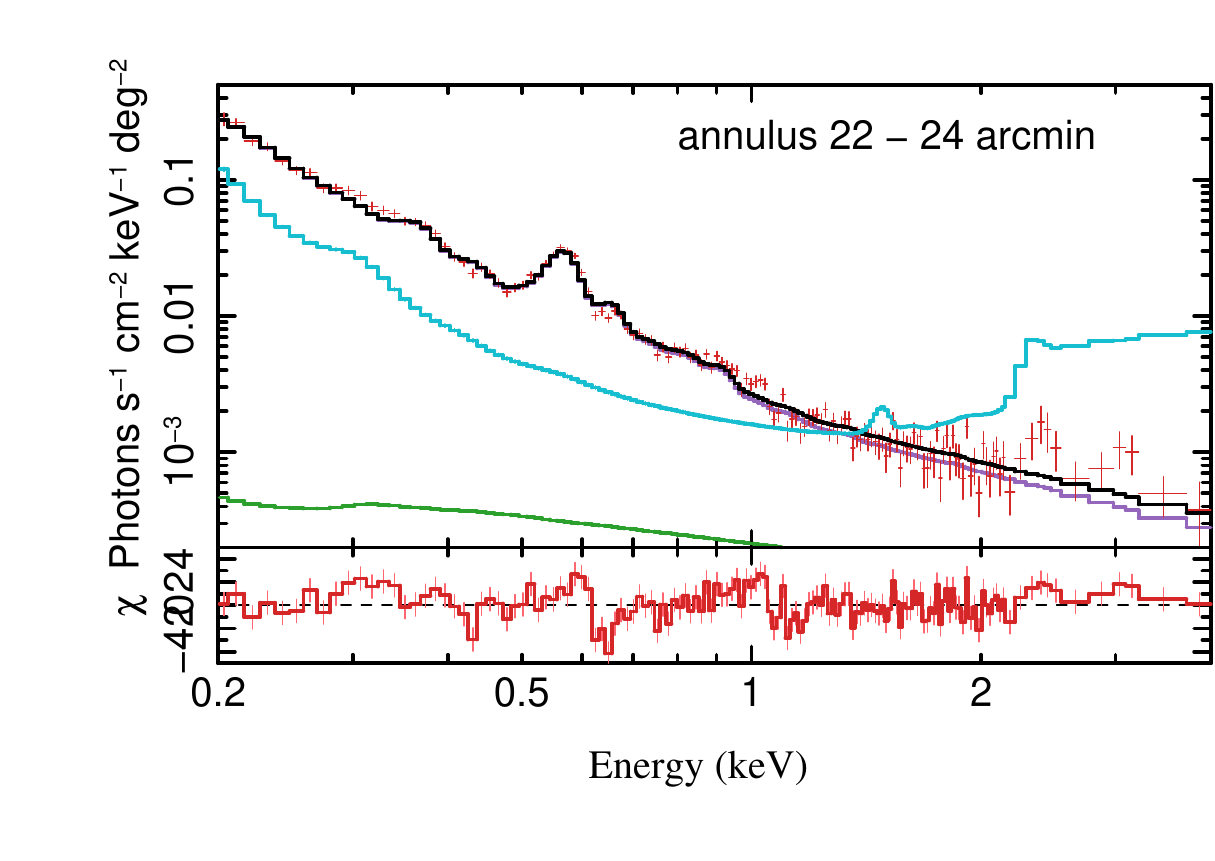}
\includegraphics[width=0.32\textwidth]{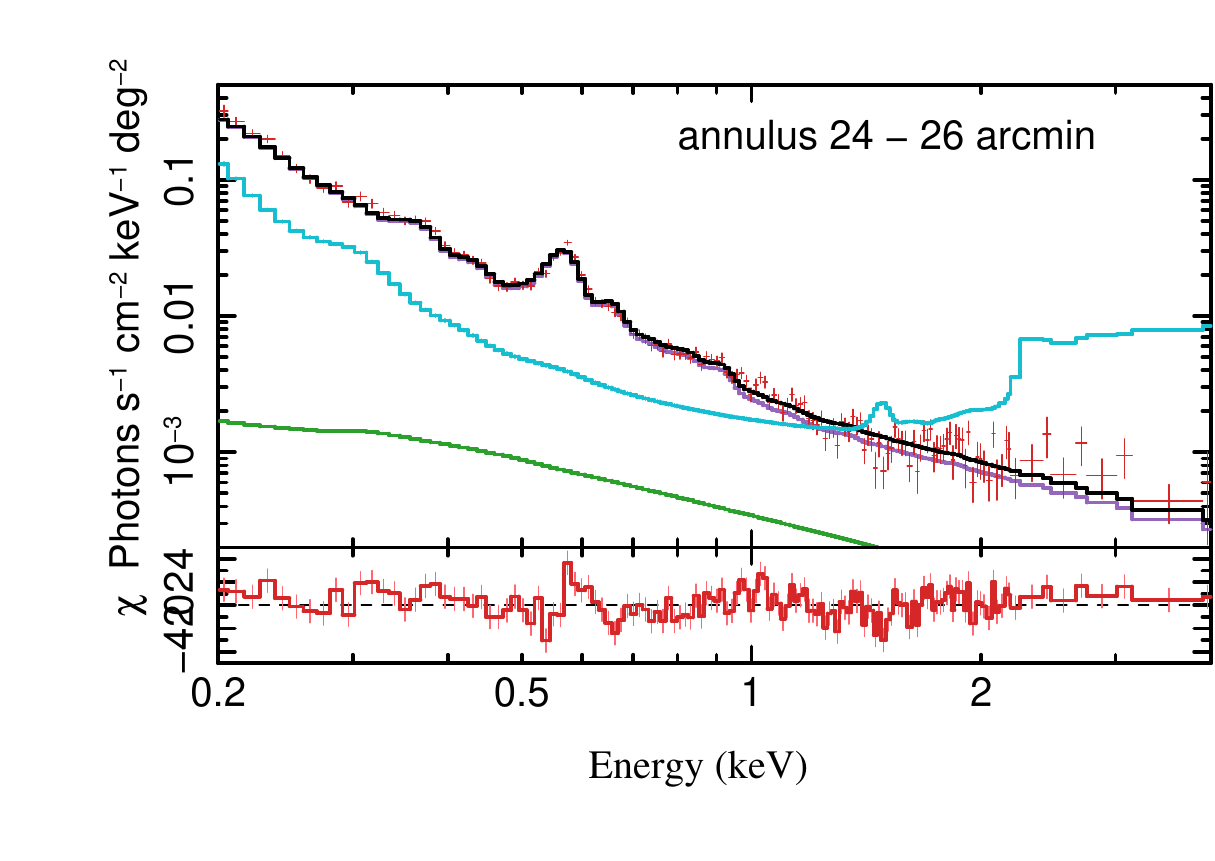}
\includegraphics[width=0.32\textwidth]{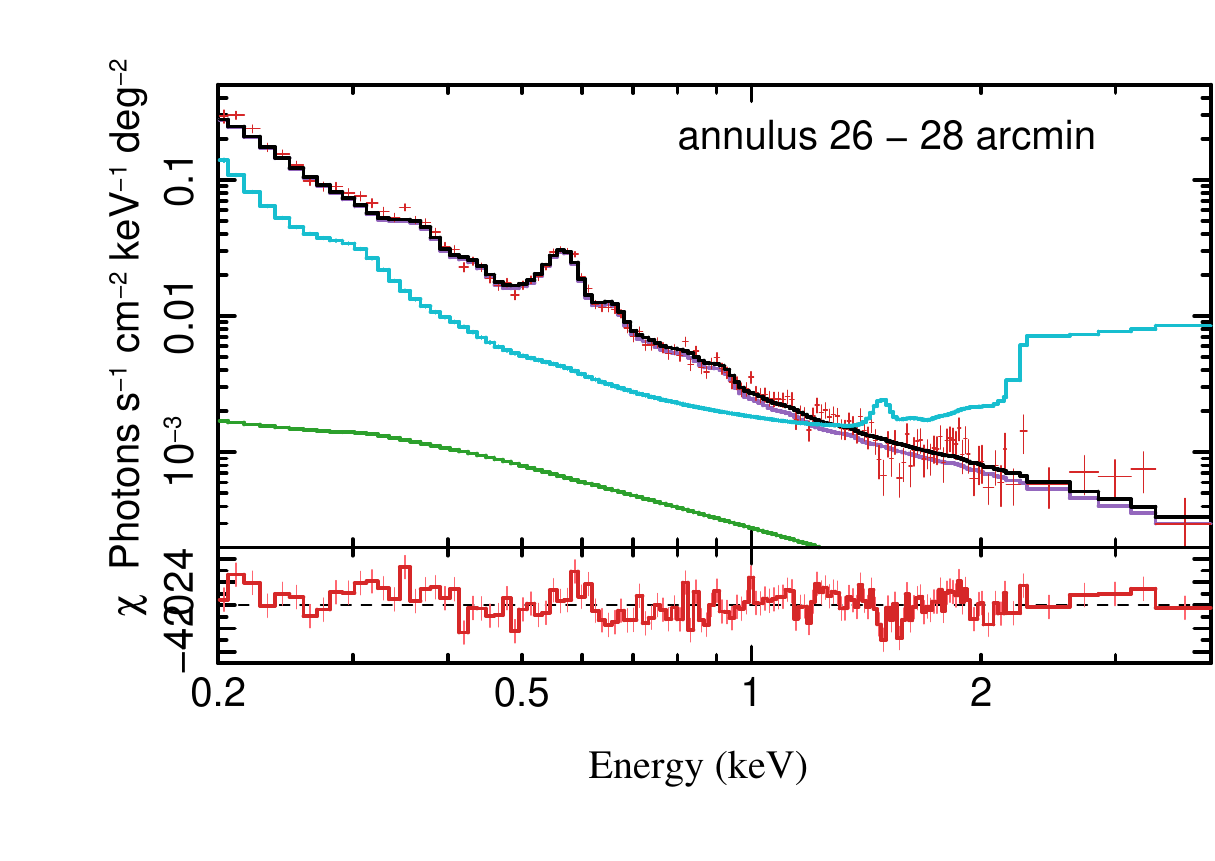}
\caption{X-ray spectrum of each annulus. The red points are the eROSITA observational data with 
the FWC data subtracted. The cyan, purple, and green lines represent the instrumental background, 
the sum of X-ray sky background, and the non-thermal excess emission. The black lines are the total 
model fitting results. Note that here the background model assumed is the physically motivated three 
APEC plus CXB model.}
\label{fig:spec}
\end{figure}

\begin{table}[!htbp]
\caption{Fitting results of the excess power-law component in each annulus, assuming the physically
motivated three APEC plus CXB background model. Fluxes are not corrected for the weak source contamination.}
\label{tab:result}
\centering
 \begin{tabular}{c | c c c c}
 \hline
 region & $\Gamma$ & Flux$_{0.2 - 2.0~\rm keV}$
 & $\chi^2$/dof \\
 (arcmin) & & ($10^{-12}~{\rm erg~cm}^{-2}~{\rm s}^{-1}~{\rm deg}^{-2}$) & \\
 \hline\hline
 4 - 6  & \multirow{12}{*}{~~$3.70^{+0.16}_{-0.15}$} 
 & $  13.28^{+1.31}_{-1.44}$ & \multirow{12}{*}{2431.286/1611}\\
 6 - 8  & & $3.78^{+0.84}_{-0.93}$ &\\
 8 - 10  & & $1.81^{+0.69}_{-0.74}$ & \\
 10 - 12 & & $0.00^{+0.49}_{-0.00}$  & \\
 12 - 14 & & $0.00^{+0.24}_{-0.00}$  & \\
 14 - 16 & & $0.41^{+0.57}_{-0.41}$  & \\
 16 - 18 & & $0.07^{+0.42}_{-0.07}$  & \\
 18 - 20 & & $0.31^{+0.31}_{-0.28}$  & \\
 20 - 22 & & $0.38^{+0.39}_{-0.33}$  & \\
 22 - 24 & & $0.73^{+0.35}_{-0.33}$  & \\
 24 - 26 & & $1.23^{+0.41}_{-0.42}$ & \\
 26 - 28 & & $1.06^{+0.48}_{-0.43}$ & \\
 \hline
 \end{tabular}
 \tablecomments{Errors are at 90\% confidence level. Columns from left to right are: annulus region, 
 photon power-law index, flux in 0.2 - 2.0 keV, and fitting $\chi^2$ value divided by the number of degree-of-freedom.}
\end{table}

\section{Validation of the PSF}
\label{app:psf}
The possible PSF leakage from the bright central source was one of the major concerns. 
To test the physical origin of the extended emission, we measured the 0.2--2.0 keV 
surface brightness profile of PSR B0656+14 after excluding all detectable point sources, 
correcting for vignetting, and normalizing to the innermost radius. We then compared its 
profile with multiple reference baselines, including both observations under the pointing 
mode and the survey mode models.

\begin{figure}[!htbp]
    \centering
    \includegraphics[width=0.6\linewidth]{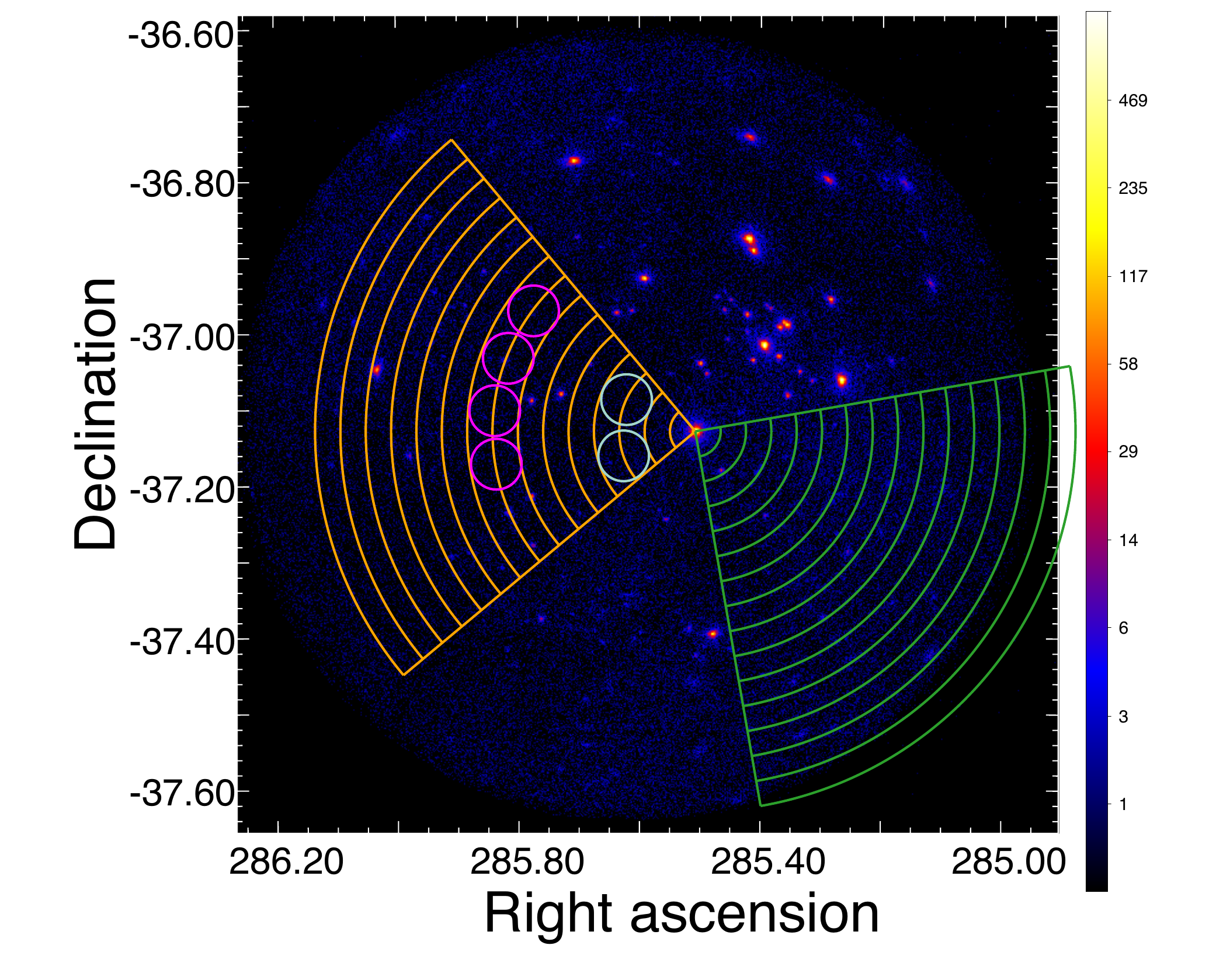}
    \caption{The eROSITA image of V* V702 CrA in 0.2--2 keV, corrected for vignetting and FWC. Two sectors show the sky regions used to extract the radial surface brightness profile, with a width of $2^{\prime}$. Small circles (in cyan and magenta) indicate the regions used to extract the spectra in Fig.~\ref{fig:v702_spec}.}
    \label{fig:v702_region}
\end{figure}

\begin{figure}[!htbp]
    \centering
    \includegraphics[width=0.6\linewidth]{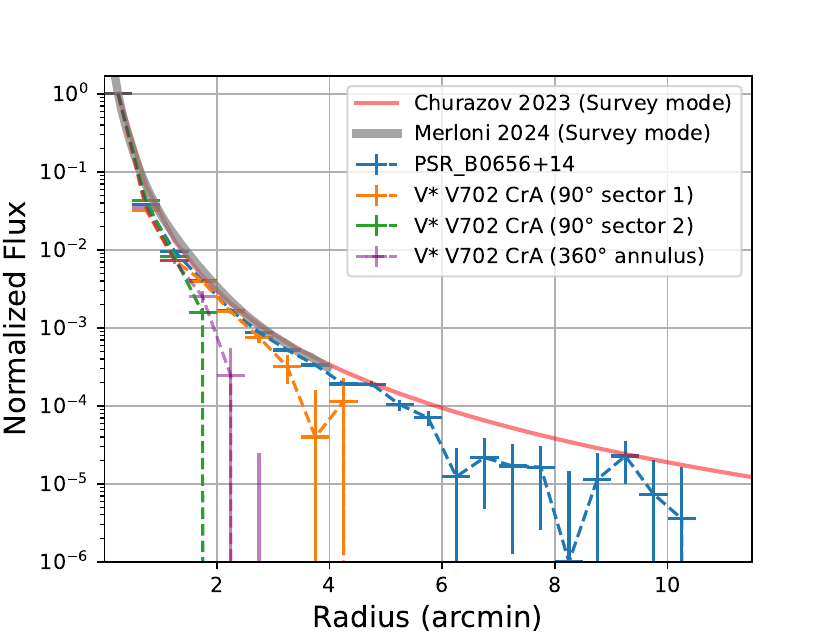}
    \caption{Comparison of the surface brightness profiles, normalized at $0.25\arcmin$. 
        Blue: PSR B0656+14 (pointing mode);  
        green and orange: V* V702 CrA (pointing mode; the 90\degr~sectors with the same color 
        as labelled in Fig.~\ref{fig:v702_region}); 
        purple: V* V702 CrA (pointing mode; 360\degr~annulus); 
        red: PSF model in the survey mode from \citet{2023A&A...670A.156C}; 
        grey: source stacking PSF in survey mode from \citet{2024A&A...682A..34M}.
        }
    \label{fig:sbp_comparison}
\end{figure}

The chosen reference source is the bright star V* V702 CrA (OBSID: 300017), which was 
observed in the same pointing mode. V* V702 CrA is a variable star of spectral type G2 
located in the field of dark cloud TGUH2213P1 \citep{2005PASJ...57S...1D}, a region with 
low background and sparse point sources. V* V702 CrA was observed at an off-axis angle of
$\sim 4.5^{\prime}$, which is slightly bigger than the off-axis angle of PSR B0656+14 
($\sim 2.5^{\prime}$). Since the eROSITA PSF naturally degrades and broadens with increasing 
off-axis angle, the reference star's PSF is expected to be intrinsically broader than that 
of the pulsar. Two reference regions of $90\degr$ sectors (see Fig.~\ref{fig:v702_region}), 
as well as the $360^{\circ}$ average, are chosen to derive the radial distributions around 
V* V702 CrA. The results are shown in Fig.~\ref{fig:sbp_comparison}. We find that the radial 
profiles for the two $90^{\circ}$ regions are slightly different, with the results of the 
green sector being more concentrated. The $360^{\circ}$ average profile lies in between of
the two selected $90^{\circ}$ sectors. This may show that there is slight asymmetry of the 
PSF for the off-axis observations. However, the current data are limited in statistics, and 
the asymmetric effect cannot be clearly characterized. In any case, the observed radial 
profile of PSR B0656+14 is more extended than that of V* V702 CrA, even if the expected 
profile should be less extended if PSR B0656+14 is indeed a point source. This implies 
that extended emission around the pulsar exists. Quantitatively, the fraction of total 
0.2--2.0 keV energy in the $4^{\prime}$--$6^{\prime}$ annulus for V* V702 CrA is 
$(0.6 \pm 0.2)\%$ as estimated using the east $90^{\circ}$ sector profile (in orange). 
For the other two profiles of V* V702 CrA, such a fraction should be even smaller.
For PSR~B0656+14, the corresponding fraction is $\sim 1.4\%$, significantly higher than 
the expectation from the V* V702 CrA profile.

For comparison, we also show in Fig.~\ref{fig:sbp_comparison} the analytical 0.4--2.3 
keV PSF model for the survey mode from \citet{2023A&A...670A.156C} (red line) and the 
official source stacking PSF of 0.2--2.3 keV \citep[grey line, retrieved from the eROSITA calibration results\footnote
{https://erosita.mpe.mpg.de/dr1/eROSITA\_technical/flight\_calibration} 
and valid only within {4'};][]{2024A&A...682A..34M}. We find that the pulsar's 
profile is apparently similar to (slightly narrower than) these survey-mode PSF profiles. 
It may partly reflect a coincidence, given the current limitations in the PSF modelling 
that the source's intrinsic extent broadens the profile to mimic the instrumental 
blurring in survey modes.

\begin{figure}[!htbp]
    \centering
    \includegraphics[width=0.6\linewidth]{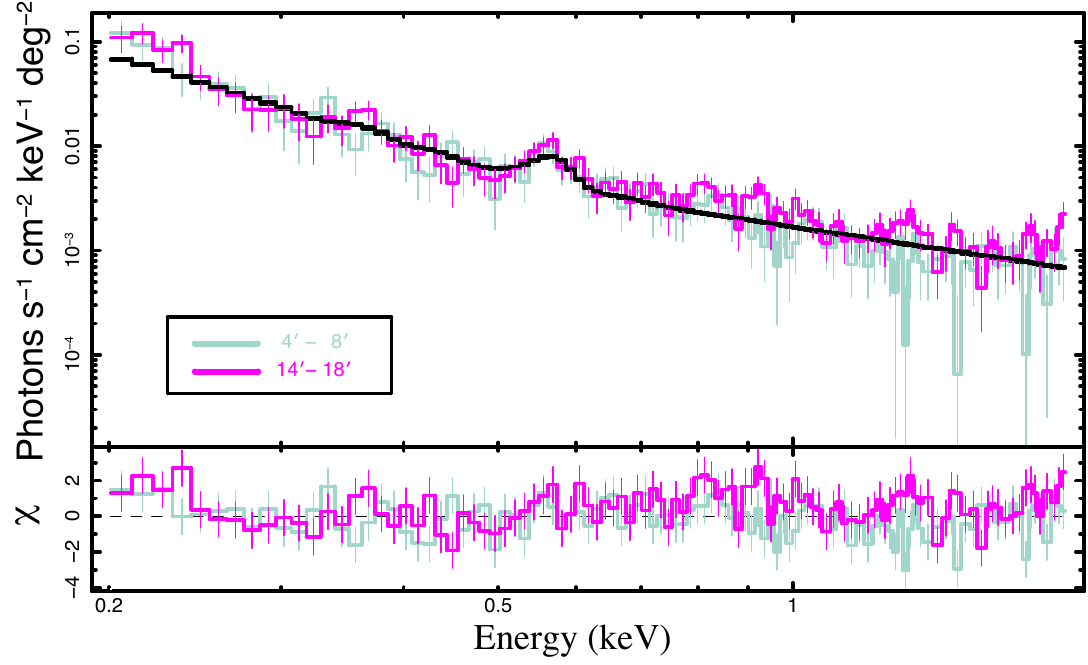}
    \caption{Simultaneous fitting of inner and outer regions around V* V702 CrA. The cyan and magenta lines correspond to the spectra extracted from the circular regions of the same colors in Fig.~\ref{fig:v702_region}, respectively.}
    \label{fig:v702_spec}
\end{figure}

Spectra were also extracted from the portions of the $4'$--$8'$ and $14'$--$18'$
radial ranges within the east $90^{\circ}$ sector regions around V* V702 CrA, as shown 
in Fig.~\ref{fig:v702_spec}. The baseline spectral model 
consists of the cosmic X-ray background plus local bubble emission with absorption, i.e.,  
{$\tt tbabs*(powerlaw+apec)$}, giving $\chi^2/\mathrm{dof} = 369.02/322$. By comparing the 
two spectra, we find no significant difference in the foreground absorption between the two 
regions, suggesting that it does not affect the surface brightness profile. This result is 
consistent with the HI observation presented by \citet{2016A&A...594A.116H}. Adding an 
additional low-metallicity {$\tt apec$} component with a fixed temperature $kT = 0.83$~keV 
(corresponding to the emission from the central V* V702 CrA) does not improve the fit 
significantly (F-test: $F = 3.466$, $p = 0.0636$).

These results indicate that the leakage of emission from the central pulsar to the
region beyond $4'$ is very limited, and the excess around PSR~B0656+14 is likely not 
due to the PSF contamination.

\section{Weak source correction}
\label{app:source_corr}
\begin{figure}[!htbp]
\centering
\includegraphics[width=0.6\textwidth]{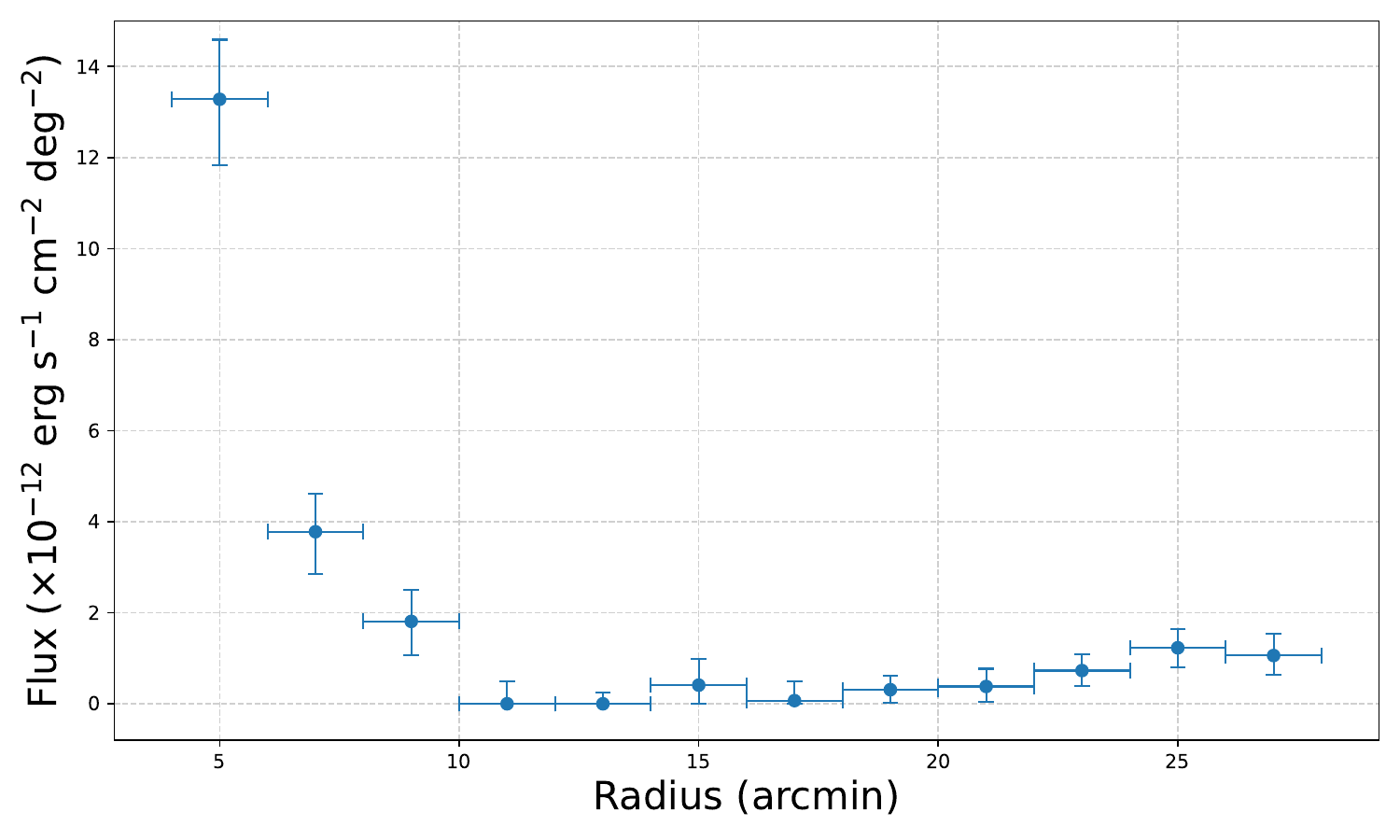}
\caption{Fluxes of excess X-rays in 0.2 - 2.0 keV band of the 12 annuli without weak source contamination correction, assuming the physically motivated three APEC plus CXB background model.}
\label{fig:flux}
\end{figure}

\begin{figure}[!htbp]
\centering
\includegraphics[width=0.6\textwidth]{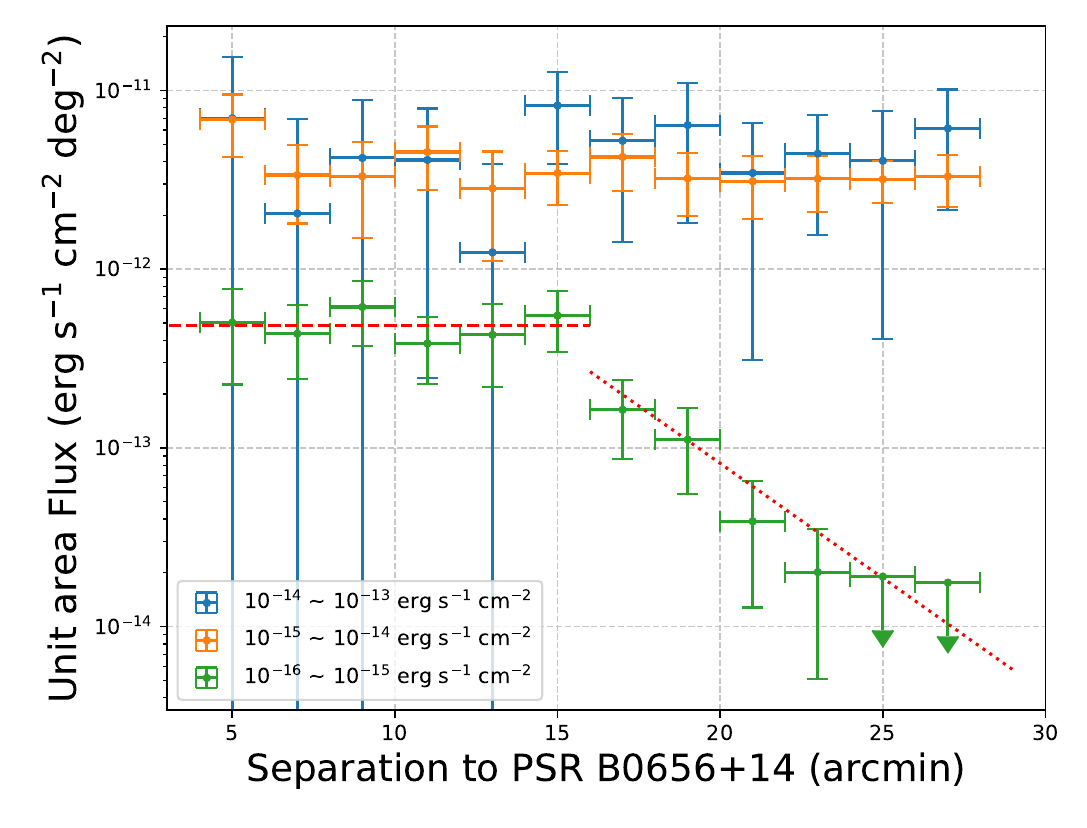}
\caption{Summed fluxes in the 0.2–2.0 keV range for three flux bins of point sources detected per unit area at different angular distances from the pulsar. The red dashed line gives the constant fitting to the low-flux results within 16$'$, and the red dotted line is the exponential fitting to the results beyond 16$'$.}
\label{fig:source_detection}
\end{figure}

The original surface brightness profile of the flux excesses is shown in Fig.~\ref{fig:flux},
which gives a rising behavior at large radii. We expect that this issue should be due to the
contamination of weak sources below the detector sensitivity.
Fig.~\ref{fig:source_detection} show the surface brightness profiles of detected sources, for 
three flux bins. We convert the count rate to the flux of these point sources with the energy 
conversion factor (ECF) and the power-law model used by \citet{2022A&A...661A...1B}. One can see 
that for relatively bright sources, the detection efficiency is almost uniform in the field of view. 
However, for the weakest sources, the detection efficiency drops significantly beyond 16$'$. 
We adopt a correction method to correct such an effect. We fit 
the surface brightness profile from 4$'$ to 16$'$ using a constant, and the surface brightness 
profile beyond 16$'$ using an exponential function. The difference between the constant surface 
brightness and the exponential function is taken as the correction values for fluxes measured 
beyond 16$'$. The corrected flux distribution is shown in Fig.~\ref{fig:flux_corrected} of the 
main text, which gives nearly flat distribution at large radii.

\section{Multi-wavelength modelling}
\label{app:sed}
\begin{figure}[!htbp]
\centering
\includegraphics[width=0.48\textwidth]{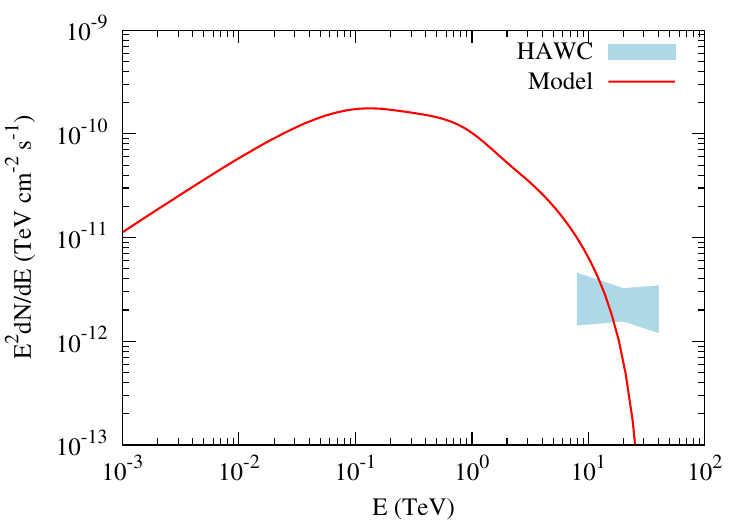}
\includegraphics[width=0.48\textwidth]{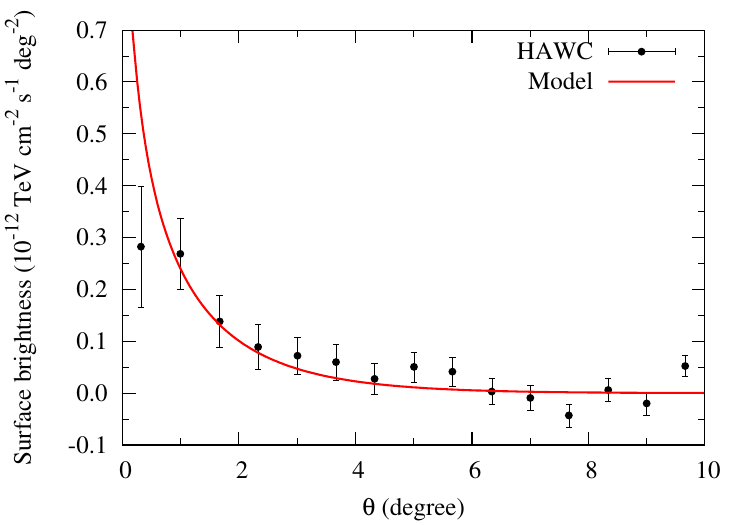}
\caption{Model predicted $\gamma$-ray spectrum (left) and surface brightness profile (right) of the non-thermal component, compared with the HAWC data \citep{2017Sci...358..911A}.}
\label{fig:model_gamma}
\end{figure}

Using the pulsar halo model described in the main text, we calculate the ICS
emission of the same population of electrons and positrons, and compare it 
with the HAWC data. The left panel of Fig.~\ref{fig:model_gamma} shows the
spectrum within 10\degr radius of the pulsar, and the right panel shows
the surface brightness profile between 8 and 40 TeV.

\bibliography{refs}

\begin{thebibliography}{}
\expandafter\ifx\csname natexlab\endcsname\relax\def\natexlab#1{#1}\fi

\bibitem[{{Abeysekara} {et~al.}(2017){Abeysekara}, {Albert}, {Alfaro},
  {Alvarez}, {{\'A}lvarez}, {Arceo}, {Arteaga-Vel{\'a}zquez}, {Avila Rojas},
  {Ayala Solares}, {Barber}, {et~al.}}]{2017Sci...358..911A}
{Abeysekara}, A.~U., {Albert}, A., {Alfaro}, R., {et~al.} 2017, Science, 358,
  911

\bibitem[{{Aguilar} {et~al.}(2016){Aguilar}, {Aisa}, {Alvino}, {Ambrosi},
  {Andeen}, {Arruda}, {Attig}, {Azzarello}, {Bachlechner}, {Barao},
  {et~al.}}]{2016PhRvL.117w1102A}
{Aguilar}, M., {Aisa}, D., {Alvino}, A., {et~al.} 2016, \prl, 117, 231102

\bibitem[{{Aharonian} {et~al.}(2021){Aharonian}, {An}, {Axikegu}, {Bai}, {Bai},
  {Bao}, {Bastieri}, {Bi}, {Bi}, {Cai}, \& et~al.}]{2021PhRvL.126x1103A}
{Aharonian}, F., {An}, Q., {Axikegu}, {et~al.} 2021, \prl, 126, 241103

\bibitem[{{Albert} {et~al.}(2023){Albert}, {Alfaro}, {Arteaga-Vel{\'a}zquez},
  {Ayala Solares}, {Belmont-Moreno}, {Capistr{\'a}n}, {Carrami{\~n}ana},
  {Casanova}, {Cotzomi}, {Couti{\~n}o De Le{\'o}n},
  {et~al.}}]{2023ApJ...944L..29A}
{Albert}, A., {Alfaro}, R., {Arteaga-Vel{\'a}zquez}, J.~C., {et~al.} 2023,
  \apjl, 944, L29

\bibitem[{{Alemanno} {et~al.}(2022){Alemanno}, {An}, {Azzarello}, {Carla
  Tiziana Barbato}, {Bernardini}, {Bi}, {Cai}, {Casilli}, {Catanzani}, {Chang},
  {et~al.}}]{2022SciBu..67.2162D}
{Alemanno}, F., {An}, Q., {Azzarello}, P., {et~al.} 2022, Science Bulletin, 67,
  2162

\bibitem[{{Berezinskii} {et~al.}(1990){Berezinskii}, {Bulanov}, {Dogiel}, \&
  {Ptuskin}}]{1990acr..book.....B}
{Berezinskii}, V.~S., {Bulanov}, S.~V., {Dogiel}, V.~A., \& {Ptuskin}, V.~S.
  1990, {Astrophysics of cosmic rays} (Amsterdam: North-Holland, 1990, edited
  by Ginzburg, V.L.)

\bibitem[{{B{\^\i}rzan} {et~al.}(2016){B{\^\i}rzan}, {Pavlov}, \&
  {Kargaltsev}}]{2016ApJ...817..129B}
{B{\^\i}rzan}, L., {Pavlov}, G.~G., \& {Kargaltsev}, O. 2016, \apj, 817, 129

\bibitem[{{Brunner} {et~al.}(2022){Brunner}, {Liu}, {Lamer}, {Georgakakis},
  {Merloni}, {Brusa}, {Bulbul}, {Dennerl}, {Friedrich}, {Liu}, {Maitra},
  {Nandra}, {Ramos-Ceja}, {Sanders}, {Stewart}, {Boller}, {Buchner}, {Clerc},
  {Comparat}, {Dwelly}, {Eckert}, {Finoguenov}, {Freyberg}, {Ghirardini},
  {Gueguen}, {Haberl}, {Kreykenbohm}, {Krumpe}, {Osterhage}, {Pacaud},
  {Predehl}, {Reiprich}, {Robrade}, {Salvato}, {Santangelo}, {Schrabback},
  {Schwope}, \& {Wilms}}]{2022A&A...661A...1B}
{Brunner}, H., {Liu}, T., {Lamer}, G., {et~al.} 2022, \aap, 661, A1

\bibitem[{{Churazov} {et~al.}(2023){Churazov}, {Khabibullin}, {Bykov},
  {Lyskova}, \& {Sunyaev}}]{2023A&A...670A.156C}
{Churazov}, E., {Khabibullin}, I., {Bykov}, A.~M., {Lyskova}, N., \& {Sunyaev},
  R. 2023, \aap, 670, A156

\bibitem[{{Di Mauro} {et~al.}(2019){Di Mauro}, {Manconi}, \&
  {Donato}}]{2019PhRvD.100l3015D}
{Di Mauro}, M., {Manconi}, S., \& {Donato}, F. 2019, \prd, 100, 123015

\bibitem[{{Dobashi} {et~al.}(2005){Dobashi}, {Uehara}, {Kandori}, {Sakurai},
  {Kaiden}, {Umemoto}, \& {Sato}}]{2005PASJ...57S...1D}
{Dobashi}, K., {Uehara}, H., {Kandori}, R., {et~al.} 2005, \pasj, 57, S1

\bibitem[{{Fang} {et~al.}(2021){Fang}, {Bi}, {Lin}, \&
  {Yuan}}]{2021ChPhL..38c9801F}
{Fang}, K., {Bi}, X.-J., {Lin}, S.-J., \& {Yuan}, Q. 2021, Chinese Physics
  Letters, 38, 039801

\bibitem[{{Fang} {et~al.}(2018){Fang}, {Bi}, {Yin}, \&
  {Yuan}}]{2018ApJ...863...30F}
{Fang}, K., {Bi}, X.-J., {Yin}, P.-F., \& {Yuan}, Q. 2018, \apj, 863, 30

\bibitem[{{H.~E.~S.~S. Collaboration} {et~al.}(2023{\natexlab{a}}){H.~E.~S.~S.
  Collaboration}, {Aharonian}, {Ait Benkhali}, {Aschersleben}, {Ashkar},
  {Backes}, {Barbosa Martins}, {Batzofin}, {Becherini}, {Berge},
  {Bernl{\"o}hr}, {Bi}, {et~al.}}]{2023A&A...673A.148H}
{H.~E.~S.~S. Collaboration}, {Aharonian}, F., {Ait Benkhali}, F., {et~al.}
  2023{\natexlab{a}}, \aap, 673, A148

\bibitem[{{H.~E.~S.~S. Collaboration} {et~al.}(2023{\natexlab{b}}){H.~E.~S.~S.
  Collaboration}, {Aharonian}, {Ait Benkhali}, {Aschersleben}, {Ashkar},
  {Backes}, {Barbosa Martins}, {Batzofin}, {Becherini}, {Berge},
  {B{\"o}ttcher}, {et~al.}}]{2023A&A...672A.103H}
---. 2023{\natexlab{b}}, \aap, 672, A103

\bibitem[{{HI4PI Collaboration} {et~al.}(2016){HI4PI Collaboration}, {Ben
  Bekhti}, {Fl{\"o}er}, {Keller}, {Kerp}, {Lenz}, {Winkel}, {Bailin},
  {Calabretta}, {Dedes}, {Ford}, {Gibson}, {Haud}, {Janowiecki}, {Kalberla},
  {Lockman}, {McClure-Griffiths}, {Murphy}, {Nakanishi}, {Pisano}, \&
  {Staveley-Smith}}]{2016A&A...594A.116H}
{HI4PI Collaboration}, {Ben Bekhti}, N., {Fl{\"o}er}, L., {et~al.} 2016, \aap,
  594, A116

\bibitem[{{Hooper} \& {Linden}(2022)}]{2022PhRvD.105j3013H}
{Hooper}, D., \& {Linden}, T. 2022, \prd, 105, 103013

\bibitem[{{Hooper} {et~al.}(2024){Hooper}, {Pinetti}, \&
  {Sokolenko}}]{2024arXiv240506739H}
{Hooper}, D., {Pinetti}, E., \& {Sokolenko}, A. 2024, arXiv e-prints,
  arXiv:2405.06739

\bibitem[{{Houck} \& {Denicola}(2000)}]{2000ASPC..216..591H}
{Houck}, J.~C., \& {Denicola}, L.~A. 2000, in Astronomical Society of the
  Pacific Conference Series, Vol. 216, Astronomical Data Analysis Software and
  Systems IX, ed. N.~{Manset}, C.~{Veillet}, \& D.~{Crabtree}, 591

\bibitem[{{Khokhriakova} {et~al.}(2024){Khokhriakova}, {Becker}, {Ponti},
  {Sasaki}, {Li}, \& {Liu}}]{2024A&A...683A.180K}
{Khokhriakova}, A., {Becker}, W., {Ponti}, G., {et~al.} 2024, \aap, 683, A180

\bibitem[{{Khokhriakova} {et~al.}(2025){Khokhriakova}, {Becker}, {Predehl},
  {Sanders}, {Freyberg}, \& {Schwope}}]{2025RNAAS...9..154K}
{Khokhriakova}, A., {Becker}, W., {Predehl}, P., {et~al.} 2025, Research Notes
  of the American Astronomical Society, 9, 154

\bibitem[{{Li} {et~al.}(2023){Li}, {Ge}, \& {Liu}}]{2023ApJ...949...90L}
{Li}, C.-M., {Ge}, C., \& {Liu}, R.-Y. 2023, \apj, 949, 90

\bibitem[{{Linden} {et~al.}(2017){Linden}, {Auchettl}, {Bramante}, {Cholis},
  {Fang}, {Hooper}, {Karwal}, \& {Li}}]{2017PhRvD..96j3016L}
{Linden}, T., {Auchettl}, K., {Bramante}, J., {et~al.} 2017, \prd, 96, 103016

\bibitem[{{Liu} {et~al.}(2019{\natexlab{a}}){Liu}, {Ge}, {Sun}, \&
  {Wang}}]{2019ApJ...875..149L}
{Liu}, R.-Y., {Ge}, C., {Sun}, X.-N., \& {Wang}, X.-Y. 2019{\natexlab{a}},
  \apj, 875, 149

\bibitem[{{Liu} {et~al.}(2019{\natexlab{b}}){Liu}, {Yan}, \&
  {Zhang}}]{2019PhRvL.123v1103L}
{Liu}, R.-Y., {Yan}, H., \& {Zhang}, H. 2019{\natexlab{b}}, \prl, 123, 221103

\bibitem[{{Ma} {et~al.}(2022){Ma}, {Bi}, {Cao}, {Chen}, {Chen}, {Cheng},
  {Gong}, {Gu}, {He}, {Hou}, {Huang}, {Huang}, {Liu}, {Shchegolev}, {Sheng},
  {Stenkin}, {Wu}, {Wu}, {Wu}, {Xiao}, {Yao}, {Zhang}, {Zhang}, \&
  {Zuo}}]{2022ChPhC..46c0001M}
{Ma}, X.-H., {Bi}, Y.-J., {Cao}, Z., {et~al.} 2022, Chinese Physics C, 46,
  030001

\bibitem[{{Manconi} {et~al.}(2024){Manconi}, {Woo}, {Shang}, {Krivonos},
  {Tang}, {Di Mauro}, {Donato}, {Mori}, \& {Hailey}}]{2024A&A...689A.326M}
{Manconi}, S., {Woo}, J., {Shang}, R.-Y., {et~al.} 2024, \aap, 689, A326

\bibitem[{{Merloni} {et~al.}(2024){Merloni}, {Lamer}, {Liu}, {Ramos-Ceja},
  {Brunner}, {Bulbul}, {Dennerl}, {Doroshenko}, {Freyberg}, {Friedrich},
  {Gatuzz}, {Georgakakis}, {Haberl}, {Igo}, {Kreykenbohm}, {Liu}, {Maitra},
  {Malyali}, {Mayer}, {Nandra}, {Predehl}, {Robrade}, {Salvato}, {Sanders},
  {Stewart}, {Tub{\'\i}n-Arenas}, {Weber}, {Wilms}, {Arcodia}, {Artis},
  {Aschersleben}, {Avakyan}, {Aydar}, {Bahar}, {Balzer}, {Becker}, {Berger},
  {Boller}, {Bornemann}, {Br{\"u}ggen}, {Brusa}, {Buchner}, {Burwitz},
  {Camilloni}, {Clerc}, {Comparat}, {Coutinho}, {Czesla}, {Dannhauer},
  {Dauner}, {Dauser}, {Dietl}, {Dolag}, {Dwelly}, {Egg}, {Ehl}, {Freund},
  {Friedrich}, {Gaida}, {Garrel}, {Ghirardini}, {Gokus}, {Gr{\"u}nwald},
  {Grandis}, {Grotova}, {Gruen}, {Gueguen}, {H{\"a}mmerich}, {Hamaus},
  {Hasinger}, {Haubner}, {Homan}, {Ider Chitham}, {Joseph}, {Joyce},
  {K{\"o}nig}, {Kaltenbrunner}, {Khokhriakova}, {Kink}, {Kirsch}, {Kluge},
  {Knies}, {Krippendorf}, {Krumpe}, {Kurpas}, {Li}, {Liu}, {Locatelli},
  {Lorenz}, {M{\"u}ller}, {Magaudda}, {Mannes}, {McCall}, {Meidinger},
  {Michailidis}, {Migkas}, {Mu{\~n}oz-Giraldo}, {Musiimenta}, {Nguyen-Dang},
  {Ni}, {Olechowska}, {Ota}, {Pacaud}, {Pasini}, {Perinati}, {Pires},
  {Pommranz}, {Ponti}, {Poppenhaeger}, {P{\"u}hlhofer}, {Rau}, {Reh},
  {Reiprich}, {Roster}, {Saeedi}, {Santangelo}, {Sasaki}, {Schmitt},
  {Schneider}, {Schrabback}, {Schuster}, {Schwope}, {Seppi}, {Serim},
  {Shreeram}, {Sokolova-Lapa}, {Starck}, {Stelzer}, {Stierhof}, {Suleimanov},
  {Tenzer}, {Traulsen}, {Tr{\"u}mper}, {Tsuge}, {Urrutia}, {Veronica},
  {Waddell}, {Willer}, {Wolf}, {Yeung}, {Zainab}, {Zangrandi}, {Zhang},
  {Zhang}, \& {Zheng}}]{2024A&A...682A..34M}
{Merloni}, A., {Lamer}, G., {Liu}, T., {et~al.} 2024, \aap, 682, A34

\bibitem[{{Ponti} {et~al.}(2023){Ponti}, {Zheng}, {Locatelli}, {Bianchi},
  {Zhang}, {Anastasopoulou}, {Comparat}, {Dennerl}, {Freyberg}, {Haberl},
  {Merloni}, {Reiprich}, {Salvato}, {Sanders}, {Sasaki}, {Strong}, \&
  {Yeung}}]{2023A&A...674A.195P}
{Ponti}, G., {Zheng}, X., {Locatelli}, N., {et~al.} 2023, \aap, 674, A195

\bibitem[{{Predehl} {et~al.}(2021){Predehl}, {Andritschke}, {Arefiev},
  {Babyshkin}, {Batanov}, {Becker}, {B{\"o}hringer}, {Bogomolov}, {Boller},
  {Borm}, {Bornemann}, {Br{\"a}uninger}, {Br{\"u}ggen}, {Brunner}, {Brusa},
  {Bulbul}, {Buntov}, {Burwitz}, {Burkert}, {Clerc}, {Churazov}, {Coutinho},
  {Dauser}, {Dennerl}, {Doroshenko}, {Eder}, {Emberger}, {Eraerds},
  {Finoguenov}, {Freyberg}, {Friedrich}, {Friedrich}, {F{\"u}rmetz},
  {Georgakakis}, {Gilfanov}, {Granato}, {Grossberger}, {Gueguen}, {Gureev},
  {Haberl}, {H{\"a}lker}, {Hartner}, {Hasinger}, {Huber}, {Ji}, {Kienlin},
  {Kink}, {Korotkov}, {Kreykenbohm}, {Lamer}, {Lomakin}, {Lapshov}, {Liu},
  {Maitra}, {Meidinger}, {Menz}, {Merloni}, {Mernik}, {Mican}, {Mohr},
  {M{\"u}ller}, {Nandra}, {Nazarov}, {Pacaud}, {Pavlinsky}, {Perinati},
  {Pfeffermann}, {Pietschner}, {Ramos-Ceja}, {Rau}, {Reiffers}, {Reiprich},
  {Robrade}, {Salvato}, {Sanders}, {Santangelo}, {Sasaki}, {Scheuerle},
  {Schmid}, {Schmitt}, {Schwope}, {Shirshakov}, {Steinmetz}, {Stewart},
  {Str{\"u}der}, {Sunyaev}, {Tenzer}, {Tiedemann}, {Tr{\"u}mper}, {Voron},
  {Weber}, {Wilms}, \& {Yaroshenko}}]{2021A&A...647A...1P}
{Predehl}, P., {Andritschke}, R., {Arefiev}, V., {et~al.} 2021, \aap, 647, A1

\bibitem[{{Profumo} {et~al.}(2018){Profumo}, {Reynoso-Cordova}, {Kaaz}, \&
  {Silverman}}]{2018PhRvD..97l3008P}
{Profumo}, S., {Reynoso-Cordova}, J., {Kaaz}, N., \& {Silverman}, M. 2018,
  \prd, 97, 123008

\bibitem[{{Schwope} {et~al.}(2022){Schwope}, {Pires}, {Kurpas}, {Doroshenko},
  {Suleimanov}, {Freyberg}, {Becker}, {Dennerl}, {Haberl}, {Lamer}, {Maitra},
  {Potekhin}, {Ramos-Ceja}, {Santangelo}, {Traulsen}, \&
  {Werner}}]{2022A&A...661A..41S}
{Schwope}, A., {Pires}, A.~M., {Kurpas}, J., {et~al.} 2022, \aap, 661, A41

\bibitem[{{Strong} {et~al.}(2007){Strong}, {Moskalenko}, \&
  {Ptuskin}}]{2007ARNPS..57..285S}
{Strong}, A.~W., {Moskalenko}, I.~V., \& {Ptuskin}, V.~S. 2007, Annu. Rev.
  Nucl. Part. Sci., 57, 285

\bibitem[{{Suzuki} {et~al.}(2025){Suzuki}, {Tsuji}, {Kanemaru}, {Shidatsu},
  {Olivera-Nieto}, {Safi-Harb}, {Kimura}, {de la Fuente}, {Casanova}, {Mori},
  {Wang}, {Kato}, {Tateishi}, {Uchiyama}, {Tanaka}, {Uchida}, {Inoue}, {Huang},
  {Lemoine-Goumard}, {Miura}, {Ogawa}, {Kobayashi}, {Done}, {Parra}, {D{\'\i}az
  Trigo}, {Mu{\~n}oz-Darias}, {Armas Padilla}, {Tomaru}, \&
  {Ueda}}]{2025ApJ...978L..20S}
{Suzuki}, H., {Tsuji}, N., {Kanemaru}, Y., {et~al.} 2025, \apjl, 978, L20

\bibitem[{{Wilms} {et~al.}(2000){Wilms}, {Allen}, \&
  {McCray}}]{2000ApJ...542..914W}
{Wilms}, J., {Allen}, A., \& {McCray}, R. 2000, \apj, 542, 914

\bibitem[{{Wu} {et~al.}(2024){Wu}, {Li}, {Liang}, {Ge}, \&
  {Liu}}]{2024ApJ...969....9W}
{Wu}, Q.-Z., {Li}, C.-M., {Liang}, X.-H., {Ge}, C., \& {Liu}, R.-Y. 2024, \apj,
  969, 9

\bibitem[{{Xi} {et~al.}(2019){Xi}, {Liu}, {Huang}, {Fang}, \&
  {Wang}}]{2019ApJ...878..104X}
{Xi}, S.-Q., {Liu}, R.-Y., {Huang}, Z.-Q., {Fang}, K., \& {Wang}, X.-Y. 2019,
  \apj, 878, 104

\bibitem[{{Yeung} {et~al.}(2024){Yeung}, {Ponti}, {Freyberg}, {Dennerl}, {Liu},
  {Locatelli}, {Mayer}, {Sanders}, {Sasaki}, {Strong}, {Zhang}, {Zheng}, \&
  {Gatuzz}}]{2024A&A...690A.399Y}
{Yeung}, M. C.~H., {Ponti}, G., {Freyberg}, M.~J., {et~al.} 2024, \aap, 690,
  A399

\end{thebibliography}
\bibliographystyle{apj}

\end{document}